\documentstyle[12pt,aaspp4]{article}
\tighten
\begin{document}
\title{Compact Group Selection From Redshift Surveys}
\author{Elizabeth Barton and Margaret J. Geller}
\affil{Harvard-Smithsonian Center for Astrophysics, 60 Garden St., Cambridge, MA 02138 \\ ebarton@cfa.harvard.edu, mgeller@cfa.harvard.edu}
\author{Massimo Ramella}
\affil{Osservatorio Astronomico di Trieste, via G.B. Tiepolo N. 11, I-34131 Trieste, Italy \\ ramella@oat.ts.astro.it}
\author{Ronald O. Marzke}
\affil{Dominion Astrophysical Observatory \\ 5071 West Saanich Road, Victoria, BC V8X 4M6, Canada \\ marzke@dao.nrc.ca}
\and
\author{L. Nicolaci da Costa\altaffilmark{1}}
\affil{European Southern Observatory \\ Karl-Schwarzschild-Strasse 2, 85748, Garching bei M\"unchen, Germany \\ ldacosta@eso.org}
\altaffiltext{1}{Departamento de Astronomia CNPq/Observat\'{o}rio Nacional} 

\begin{abstract}

For the first time, we construct a catalog of compact groups
selected from a complete, magnitude-limited redshift survey.  
We select groups with $N \geq 3$ members based on projected separation
and association in redshift space alone.  
We evaluate the characteristics of the Redshift Survey Compact Groups (RSCG's).
Their physical properties (membership frequency,
velocity dispersion, density) are similar to those of the Hickson 
[ApJ, 255, 382 (1982)]
Compact Groups.  Hickson's isolation criterion is a strong
function of the physical and angular 
group radii and is a poor predictor of the group environment.
In fact, most RSCG's are embedded in dense environments.
The luminosity function for RSCG's is mildly inconsistent with the
survey luminosity function --- the characteristic luminosity is brighter and
the faint end shallower for the RSCG galaxies.
We construct a model of the selection function of compact groups.  Using
this selection function, we estimate the abundance of RSCG's;
for groups with $N \geq 4$ members the abundance is
$3.8 \times 10^{-5}\ {h}^3\ {\rm Mpc}^{-3}$.  For all
RSCG's ($N \geq 3$) the abundance is
$1.4 \times 10^{-4}\ {h}^3\ {\rm Mpc}^{-3}$.

\end{abstract}

\section{Introduction}

Compact groups are the densest known systems of galaxies in the universe.  
Rose \markcite{r77} (1977) and Hickson 
\markcite{h82} (1982) made the first large-scale, systematic searches
for dense systems on the sky.
Subsequent studies of the properties and environments of Hickson's 100
Compact Groups indicate that they are probably not dynamically simple,
isolated systems.  

Here, for the first time, we
select compact systems from a complete, magnitude-limited redshift
survey.  
We select groups based on physical extent, rather
than angular size.  
This approach eliminates some of the systematic biases intrinsic to
identification of systems on the sky.  Because we begin with a complete,
magnitude-limited survey we can model the residual selection effects
from first principles.
We use the redshifts of galaxies surrounding our
compact groups to explore the embedding of the 
Redshift Survey Compact Groups (RSCG's, hereafter) in their environments.

Hickson's work
sparked debates about the physics of compact groups.
The existence of compact groups is a challenge 
for dynamical models because their measured crossing times are generally much 
smaller than the Hubble time.  Numerical simulations 
indicate that compact group galaxies merge to form
bright elliptical galaxies on timescales comparable to the group crossing times
(Barnes \markcite{b89}  1989; Carnevali, Cavaliere \&
Santangelo \markcite{c81} 1981; Cavaliere et al.\ \markcite{c83} 1983; 
Governato, Bhatia \& Chincarini \markcite{b91} 1991).
To resolve this problem, Mamon \markcite{m86} (1986) suggested that
a large fraction of compact groups are merely unbound chance superpositions
of galaxies within loose groups (Mamon \markcite{m86}  1986, 
\markcite{m87} 1987).  Hernquist,
Katz \& Weinberg \markcite{hkw95} (1995) suggested that compact groups are
unbound superpositions of galaxies viewed along filaments.  In contrast,
simulations by Hickson \& Rood \markcite{h88} (1988)
and  Diaferio, Geller \& Ramella 
\markcite{d95} (1995) and
observational work by Zepf \markcite{z93} (1993) and Pildis, Bregman \&
Schombert \markcite{p95} (1995) suggest that many of Hickson's compact groups 
(HCG's, hereafter)
are, in fact, bound systems.

The short crossing times of HCG's
are not a problem if the groups
form continually in dense environments like loose groups
(Barnes \markcite{b89} 1989, Diaferio, Geller \& Ramella 
\markcite{d94} 1994).  Studies of the environments around
some of the HCG's reveal that many of them are indeed embedded in larger,
looser systems (Ramella et al.\ \markcite{r94} 1994; 
de Carvalho, Ribeiro \& Zepf \markcite{c94} 1994).  

Compact groups have implications for cosmology, 
the development of large-scale structure and the evolution of the
galaxy population. Compact groups are more likely to form
during the present epoch in a dense universe (Diaferio \markcite{ad94}  1994, 
Governato, 
Tozzi \& Cavaliere \markcite{g95} 1995).  
Diaferio, Geller \& Ramella \markcite{d94} (1994)
and Ramella et al.\ \markcite{r94} (1994) realized that compact groups may 
be a clue to the evolutionary state of loose groups.
If compact groups are dense environments, galaxy interactions and mergers
within them are likely (Barnes \markcite{b89} 1989).  

Here, we apply an objective group-finding algorithm for identifying
HCG-like systems in the CfA2+SSRS2 Redshift Survey.
In \S 2 we describe previous sky-selected compact group surveys.  In
\S 3 we describe the redshift surveys.  \S 4 contains a description
of our group-finding algorithm and the biases it introduces.  
In \S 5 we present the catalog and evaluate the physical properties
of RSCG's. In \S 6 we apply Hickson's (1982) selection criteria 
to RSCG's.  \S 7 contains an analysis of the environments of
RSCG's.  In \S 8 we evaluate the luminosity function of
galaxies in RSCG's.  We then calculate the selection function
and compute the resulting abundance of RSCG's.

\section{Compact Group Selection on the Sky}

Examination of photographic plates led to the identification of 
compact groups as unusually dense, apparently isolated knots of
galaxies on the sky (Shakhbazyan \markcite{s73} 1973; 
Vorontsov-Velyaminov \markcite{v59} 1959, \markcite{v77} 1977;
Burbidge \& Burbidge \markcite{b61} 1961; 
Arp \markcite{a66} 1966).  Rose \markcite{r77} (1977) made the first large
statistical study of these systems.  Hickson
(1982) later defined quantitative criteria for population, isolation on the sky,
and compactness.  He identified 100 HCG's
on the POSS R plates
which satisfy 3 criteria:

\begin{itemize}
\item $N \geq 4$ where $N$ is the number of members within 3 magnitudes
of the brightest galaxy, $m_{1}$.
\item $\theta_N \geq 3 \theta_G$ where $\theta_{G}$ is the angular radius
of the smallest circle which contains the geometric centers of all the 
suggested group members.  The radius $\theta_N$ is the angular radius 
of the largest concentric circle which contains no further galaxies with
$m < m_1 + 3$.
\item $\mu_G < 26$ where $\mu_G$ is the total magnitude of the galaxies
in the R band averaged over the circle of radius $\theta_G$, 
in magnitudes~arcsecond$^{-2}$.
\end{itemize}

Hickson examined the entire POSS including regions at low Galactic
latitude.  Later, Hickson
et al.\ \markcite{h92} (1992) measured radial velocities of galaxies in HCG's, 
and defined a set of 92 HCG's with $N \geq 3$
where $N$ now refers to the number of members within 1000~km~s$^{-1}$ of
the median group velocity.  The median redshift of these HCG's 
is 0.030.
Prandoni, Iovino \& MacGillivray \markcite{p94} (1994) 
applied Hickson's original
criteria to automated plate scans and identified 59 candidate compact systems. 

These catalogs of systems on the sky have a number of unavoidable 
systematic problems.  First, in spite of their large projected surface
density, the group candidates may include interlopers with redshifts
substantially different from the
systemic mean.  Among the 100 systems originally identified by Hickson,
92\% have $N \geq 3$; only 69\% actually have $N \geq 4$ 
(Hickson et al.\ \markcite{h92} 1992).  
Selection on the sky unavoidably introduces correlations of group
properties with distance.  For example, nearby systems with a large
angular scale are underrepresented in these catalogs.

\section{The CfA2 + SSRS2 Redshift Survey}

We select compact groups from redshift surveys which include 14,383 
galaxies.
CfAnorth covers the declination range $8.5^\circ \leq \delta \leq 44.5^\circ$
and right ascension range $8^{h} \leq \alpha \leq 17^{h}$ (B1950) and includes 6500
galaxies (Geller \& Huchra \markcite{g89} 1989; 
Huchra et al.\ \markcite{h90} 1990; Huchra et al.\ \markcite{h95} 1995). 
CfAsouth covers the 
region $-2.5^\circ \leq \delta \leq 48^\circ$ and 
$20^{h} \leq \alpha \leq 4^{h}$ and includes 4283 galaxies
(Giovanelli \& Haynes \markcite{g85} 1985; Giovanelli et al.\ 
\markcite{g86} 1986;
Haynes et al.\ \markcite{hay88} 1988; Giovanelli \& Haynes \markcite{gh89} 1989; 
Wegner, Haynes \& Giovanelli \markcite{w93} 1993; 
Giovanelli \& Haynes \markcite{g93} 1993; Vogeley \markcite{v93} 1993).  
SSRS2 includes 3600 
galaxies and is complete over
1.13 steradians of the of the southern galactic cap in the declination
range $-40^\circ \leq \delta \leq -2.5 ^\circ$ and $b \leq -40^\circ$
(da Costa et al.\ \markcite{da94} 1994).  
Both SSRS2 and CfA2 are magnitude-limited to
$m_{B_{0}} \leq 15.5$.  SSRS2 is derived from plate scans.  CfA2 is based on
the Zwicky catalog.  We use only the $cz$ range $300\ {\rm km\ s}^{-1}$
to $15,000\ {\rm km\ s}^{-1}$; the full sample we 
examine includes 14,011  galaxies.

The coordinate uncertainties in the redshift catalog are about
one arcminute. 
We test the possible effects of the uncertainties on the very dense RSCG's.
They do not affect our compact group catalog.

\section{Compact Group Selection in Redshift Space}

We develop an algorithm for identifying compact systems from a complete 
redshift survey.  Our criteria mimic the ones established by Hickson.  
Application of our procedure to a {\it distance limited} redshift
catalog would yield a set of systems with properties independent of distance.
To obtain the largest possible sample of systems, we analyze
a {\it magnitude limited} redshift survey.  This limitation introduces
some biases as a function of distance, but
they are less severe than those
introduced by selection on the sky.  Because we start with a complete 
survey we can account for these biases.

The objective algorithm we use to select compact group candidates from the
redshift surveys is a modification of the friends-of-friends 
algorithm developed by Huchra \& Geller \markcite{hg82} (1982).  
We use a new group-finding code from Ramella, Pisani \& Geller \markcite{r296} (1996).
We identify groups of galaxies as linked sets
of ``neighboring'' galaxies.  To determine whether two
galaxies belong to a group, we  consider both their projected
separation, $\Delta D$, and their line-of-sight velocity difference,
$\Delta V$.
The projected separation of the pair is 
$\Delta D=2\left(\frac{v}{H_{0}}\right)\sin\left(\frac{\Delta \theta}{2}
\right)$,
where $\Delta \theta$ is the angular 
separation on the sky and $v=cz$ is the average 
redshift.  Throughout the paper we use 
$H_{0} = 100\ {\rm km \  s}^{-1}{\rm Mpc}^{-1}$.

We restrict the size of our groups by specifying limiting parameters,
$D_{0}$ and $V_{0}$.  If $\Delta D \leq D_{0}$ 
and $\Delta V \leq V_{0}$, the galaxies are neighbors. 
We search each galaxy in a pair of neighbors for additional neighbors.  
Linked sets of neighbors are groups.
Groups with three or more
members constitute our objective sample of compact
groups.  We select values of $V_{0}$ and $D_{0}$ which produce a catalog
of systems with properties similar to those of Hickson Compact Groups.
We keep $D_{0}$ fixed to ensure that we identify only systems where inter-galaxy
separations are comparable with their physical size.

The line-of-sight velocity difference between galaxies in a gravitationally
bound system is a measure of their relative peculiar velocity.
The median radial
velocity dispersion in HCG's is only 
$\sim$ 200~km~s$^{-1}$ (Hickson et al.\ \markcite{h92} 1992).  
We choose the value $V_{0}$ = 1000 ~km~s$^{-1}$,
which is large enough to include most
physically associated galaxies in compact groups.  
This value of $V_{0}$ is, however, not so large that the
resulting ``groups'' accidentally span 
voids in the galaxy distribution (Geller \& Huchra \markcite{g89} 1989).
This value is consonant with Hickson's 
procedure of rejecting galaxies
with velocities different from the median
group velocity by $\geq$ 1000~km~s$^{-1}$ (Hickson et al.\ \markcite{h92} 1992).

\begin{figure}[p]
\centerline{\epsfxsize=3.7in%
\epsffile{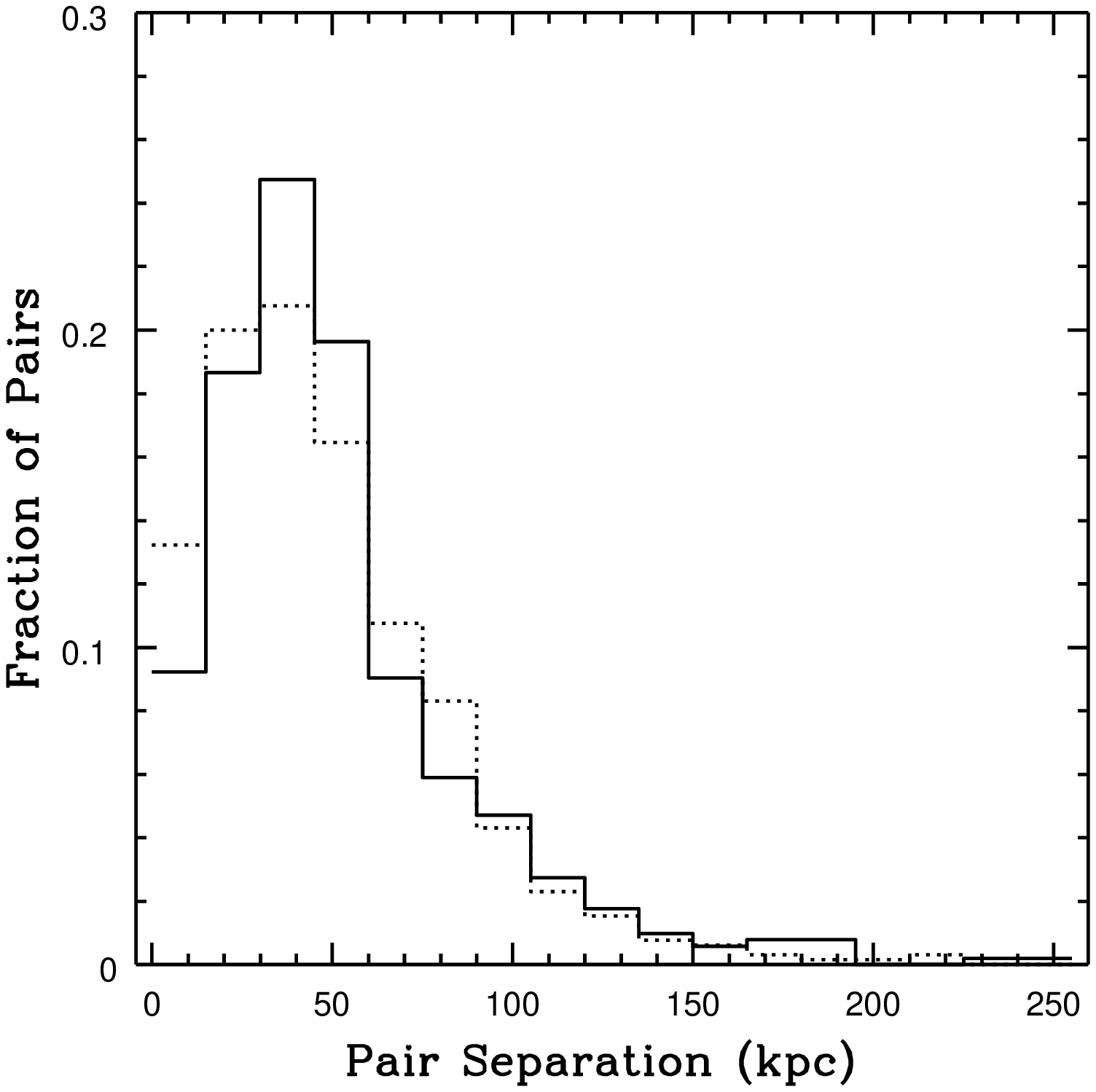}}
\caption{Projected separation distributions of the CfAnorth
(solid line) and Hickson Compact Group Catalogs (dotted line).} \label{fig:pairsep}

\centerline{\epsfxsize=3.7in%
\epsffile{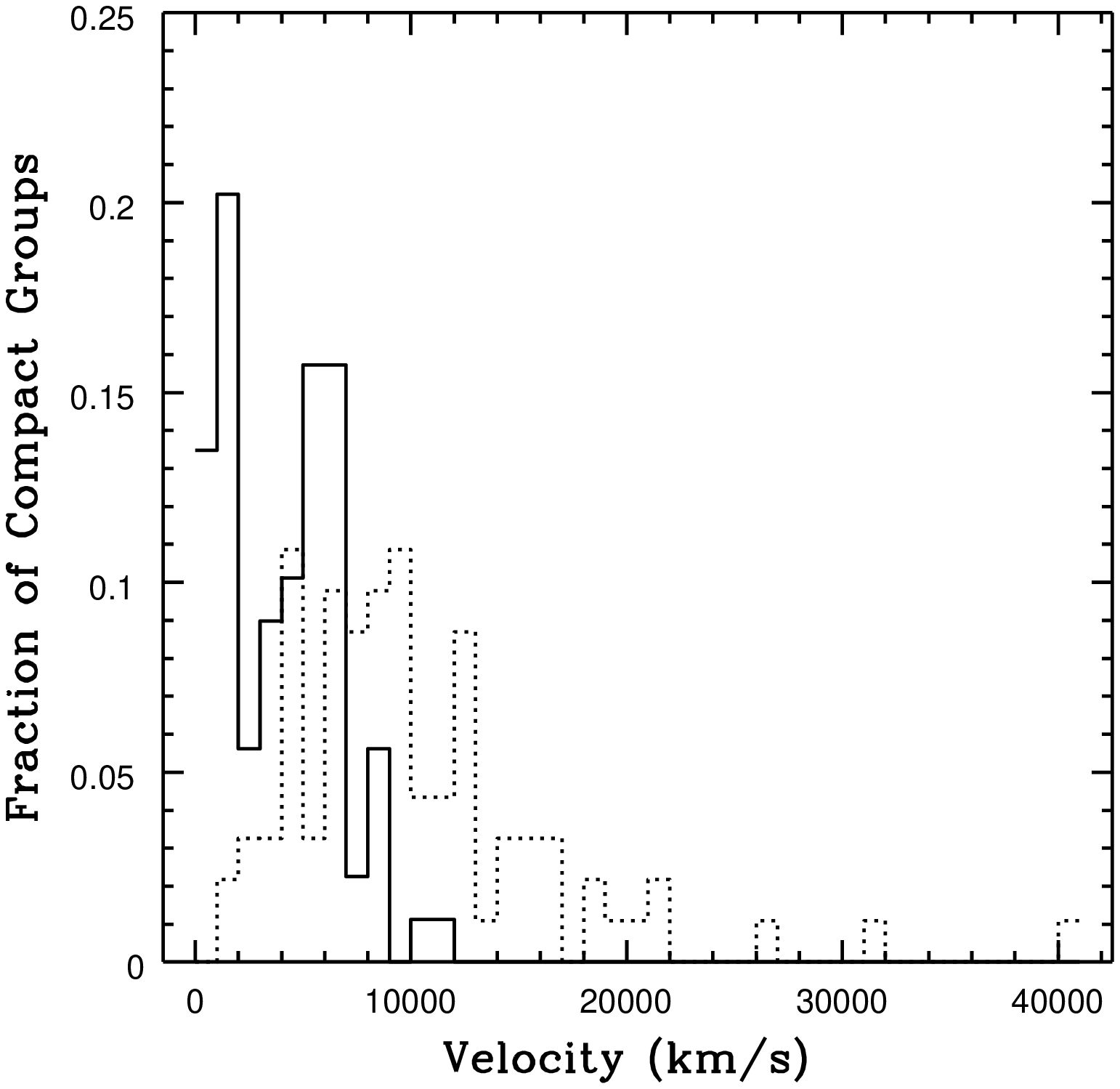}}
\caption{Redshift distribution of HCG's (dotted line) and
RSCG's (solid line).} \label{fig:n_of_z}
\end{figure}

The parameter $D_{0}$ directly limits the physical extent of
our compact groups.  We 
use $D_{0}$ to match our sample to the HCG's.
As a measure of the spatial extent of HCG's, we
use the distribution of all projected 
separations, $\Delta D$, among Hickson group members.  
We compare the distribution to the distribution of all
projected separations in compact group catalogs extracted from the CfAnorth 
catalog.
We apply the friends-of-friends algorithm 
to construct catalogs of compact group candidates 
at several values of $D_{0}$.  
Figure~\ref{fig:pairsep} shows that the value $D_{0}$ = 50 kpc yields the best match between
CfAnorth (solid line) and the HCG's (dashed line).
A K-S test indicates that the null
hypothesis that the density values are drawn from the same distribution
is acceptable at the 19\% confidence level.
On the basis of this match, the catalog with
$V_{0}$~=~1000~km~s$^{-1}$ and $D_{0}$~=~50~kpc  is
our objectively selected catalog of compact groups, the RSCG's.
Table~\ref{tab:sepdist} lists the median, first and third 
quartiles of the separation distributions
for the HCG catalog and the CfAnorth catalog of compact groups.  
Table~\ref{tab:sepdist} also includes physical parameters 
for the RSCG's we extract from CfAsouth, SSRS2 and CfA2+SSRS2 using
$D_{0}=50$~kpc and $V_{0}=1000$~km~s$^{-1}$.
Note that the third quartile value for the SSRS2 sample (and for
the CfA + SSRS2 sample) is inflated by the presence of a single group of 13 
galaxies.

The physical properties of RSCG's are 
similar to the HCG's,  even though we do not 
implement all of Hickson's selection criteria directly.
First, we do not take galaxy magnitudes into account in the group selection; 
some of our systems have fewer than three galaxies
in the interval [$m_{1}$,$m_{1}+3$].  
Furthermore,
many of our groups have $m_{1} + 3 > m_{\rm lim}$, 
where $m_{\rm lim,Zw}=15.5$ is the
limiting magnitude of the survey. Thus we do not necessarily
include all of the galaxies which might be in the interval [$m_{1}$,$m_{1}+3$].
We later examine POSS images of the groups and find few fainter
galaxies inside the group radii.

Second, we do not implement Hickson's isolation
criteria in our initial group selection.  
Third, we do not reject groups on the
basis of surface brightness. However, all but four very nearby groups in
our catalog automatically satisfy the surface brightness criterion because
we require galaxy projected separations comparable with the size of a galaxy.
Finally, we note that unlike the HCG's, our sample is
a complete listing of compact groups of three or more galaxies.  Hickson 
includes a triplet only when one member of his initial group has a
discordant redshift.

In general, our limits on the physical extent of compact groups are
more restrictive than Hickson's.
23 HCG's are within the redshift survey boundaries and contain at least
three galaxies brighter than the magnitude limit.  We detect parts or all
of 15 of these HCG's.  We fail to detect the other HCG's because the physical
or velocity separation of member galaxies exceeds our limits, or because of
magnitude errors.  

Both our selection procedure and Hickson's introduce biases as a function
of redshift or distance.  The most important 
bias is the dependence of density on
distance.  At large distances, any magnitude-limited survey samples only the
bright end of the luminosity function.  Both our sample of 
galaxies and Hickson's are magnitude-limited;  Hickson's limit is fainter.
Because our selection criteria do not vary with redshift to 
compensate for the magnitude limit, the systems we detect at large
distances are generally denser than the average nearby system.  
Hickson's sample suffers from a similar bias.
His distant groups are more likely to have larger total populations
because of the magnitude limit, but there is no fixed upper limit
to the physical inter-galaxy separations.  
We model these effects in Section~8.2.

Nearby groups with large angular radii are absent from the HCG catalog
because of the
large probability of interlopers within the annulus
$\theta_{G} \leq \theta \leq 3 \theta_{G}$.  These nearby groups are also
more difficult to spot by eye.  
Figure~\ref{fig:n_of_z}
shows the redshift distribution of HCG's (dotted line) and 
RSCG's (solid line).  We identify groups where Hickson's approach
is least effective; the two compact group surveys complement one another.

\section{The Compact Group Catalog}

The search algorithm, applied to the CfA2+SSRS2 survey,
yields a catalog of 89 groups of three or more galaxies with properties
similar to the HCG's.  
There are 50 groups in CfAnorth, 23 groups in CfAsouth and 16 groups
in the SSRS.  
15 of these groups are HCG's or subsets of HCG's.  
Table~\ref{tab:prop} lists the locations and basic properties of these
groups.  The median redshift
of our sample of compact groups is $z=0.014$, only half of
the median HCG redshift, $z=0.030$.  

\begin{figure}[p]
\caption{RSCG 7:  The inner circle is the smallest circle 
(radius $\theta_G = 2.99$~arcmin) enclosing all the
centers of the galaxies in the RSCG.  The outer circle is 
$3 \theta_G$.  We calculate the circle with coordinates derived
from the Digitized Sky Survey for this figure. The figure is available
via anonymous ftp at ftp://cfa0.harvard.edu/outgoing/barton.} \label{fig:rscg7}

\vspace{50pt}

\caption{RSCG 47:  The inner circle is the smallest circle 
(radius $\theta_G = 2.01$~arcmin) enclosing all the
centers of the galaxies in the RSCG.  The outer circle is 
$3 \theta_G$.  We calculate the circle with coordinates derived
from the Digitized Sky Survey for this figure. The figure is 
available via anonymous ftp at ftp://cfa0.harvard.edu/outgoing/barton.} \label{fig:rscg47}

\vspace{50pt}

\plottwo{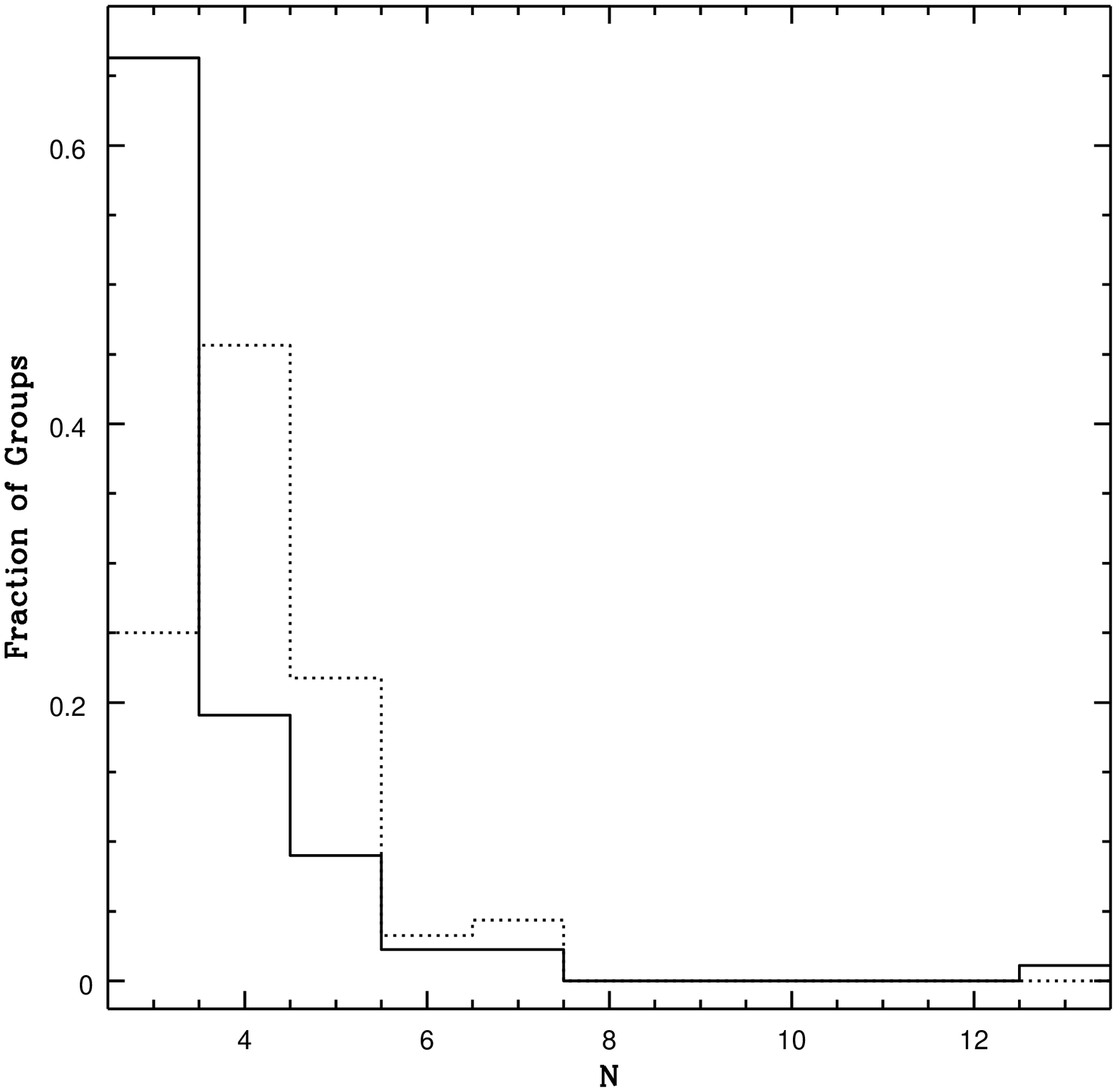}{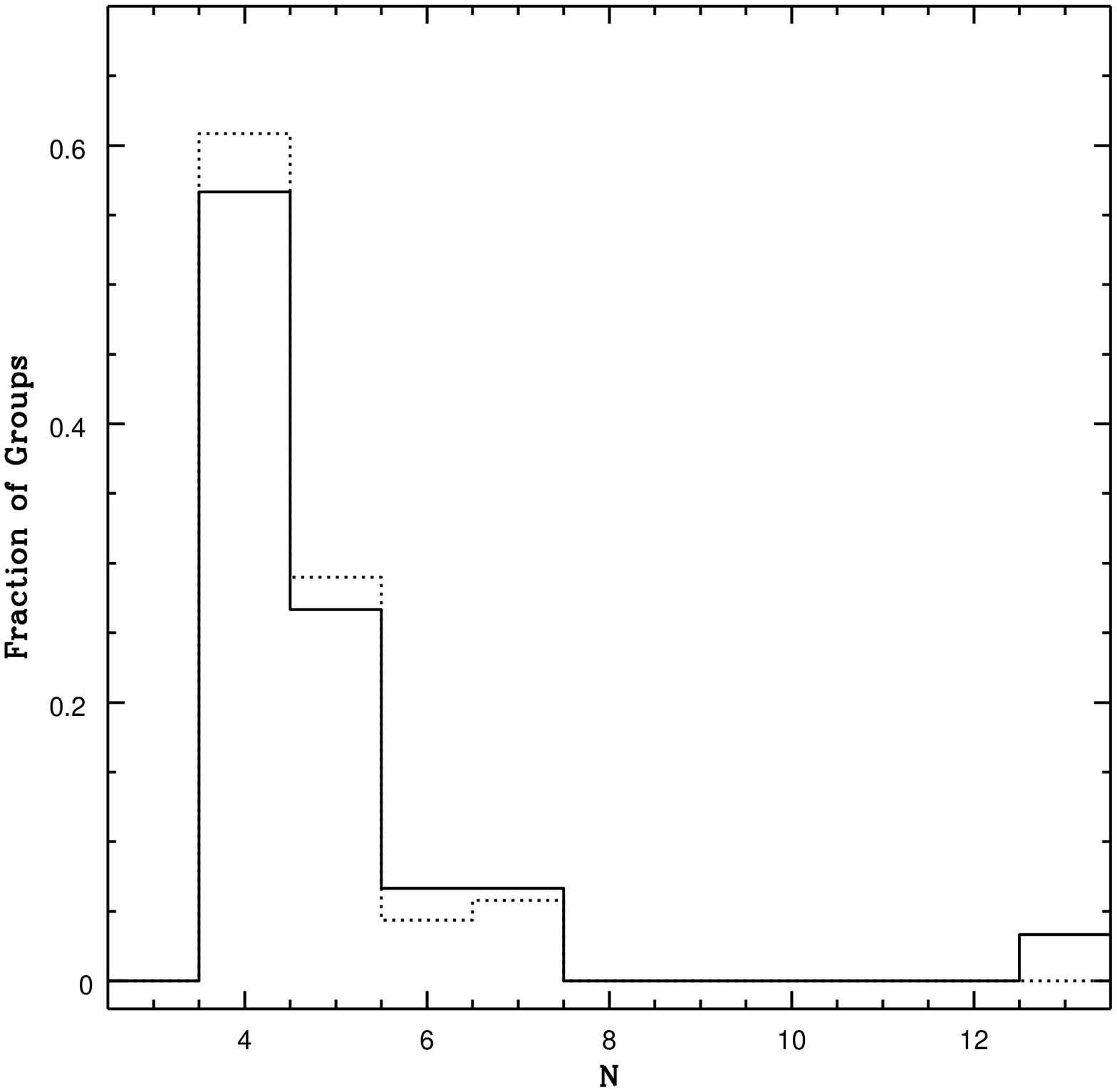}
\caption{Population frequency distributions of RSCG's and HCG's: (a) RSCG's with
$N \geq 3$ (solid line) and HCG's with $N \geq 3$ (dotted line), and 
(b) RSCG's with $N \geq 4$ (solid line) and
HCG's with $N \geq 4$ (dotted line). } \label{fig:freq}
\end{figure}

Figures~\ref{fig:rscg7}~and~\ref{fig:rscg47} show Digitized Sky Survey
images of RSCG~7 and RSCG~47, respectively.  Neither RSCG is an HCG.
In each image, the inner circle
is the smallest circle containing the centers of all member galaxies.  It 
has angular radius $\theta_G = 2.94$~arcmin for RSCG~7 and $\theta_G=
2.14$~acrmin for RSCG~47.  The outer circle has radius $3\theta_G$.  
RSCG~7, a group with $N = 3$, has galaxies in its isolation annulus.  
The two brightest galaxies
in this annulus are in the survey, but their velocities are about 
1600~km~s$^{-1}$ from the group median.  The other galaxies are too faint to
be in the survey.  RSCG~47, a group with $N = 4$ galaxies, is isolated
according to Hickson's criterion.

Because of the large scale structure in the redshift
surveys, the distribution of groups with redshift is nonuniform and
varies among the three portions of the survey.  The 
median group redshifts for individual portions of the
survey are:  
$z=0.007$ for CfAnorth,
$z=0.017 $ for CfAsouth and $z=0.012$ for SSRS2.  

Next, 
we explore the properties of the RSCG's and compare them with the HCG's.
Although the selection algorithms for HCG's and RSCG's are not 
identical, the systems in the two catalogs have similar physical 
properties.

\subsection{Membership Frequency}
Figure~\ref{fig:freq}a shows the frequency of occurrence
of groups of all populations $N$ in our catalog 
(solid line) and in the HCG catalog
(dashed line).  The overall distributions do not match.  
Hickson only finds triplets when his groups selected on the
sky contain interlopers;  in other words, triplets are underrepresented
relative to our sample.
However, the match is 
excellent for groups with $N \geq 4$ (Figure~\ref{fig:freq}b), when the
distribution is renormalized.  
Both distributions also agree with models of compact groups formed as
subsystems within
loose groups (Diaferio, Geller \& Ramella \markcite{d94} 1994).
In this section, we consider the distributions of various 
compact group properties separately for groups of $N= 3$ and 
groups of $N \geq 4$.  One physical motivation is the much greater 
likelihood that tight groups of $N \geq 4$ are bound.
Diaferio, Geller \& Ramella \markcite{d94} (1994) 
find that within loose groups
only 50\% of triplets are bound systems,
whereas 80\% of systems with four or more galaxies are bound systems.  

\subsection{Velocity Dispersion}
In Figures~\ref{fig:veldisp}a and b, we plot the distribution of the
velocity dispersions of the HCG's  (dashed line) 
and of the 89 RSCG's (solid line).  Figure~\ref{fig:veldisp}a is 
the distribution
for groups of 3 and Figure~\ref{fig:veldisp}b includes only 
groups of $N \geq 4$. 
Table~\ref{tab:veldisp} lists the median, first and third quartiles
of these distributions. 
The distributions are similar.  K-S tests
indicate that the probability these samples are
drawn from the same parent distribution is 
$86.5\% $ for groups with $N=3$
and $26.5\% $ for groups with
$N \geq 4$.

\subsection{Density} 
\begin{figure}[p]
\plottwo{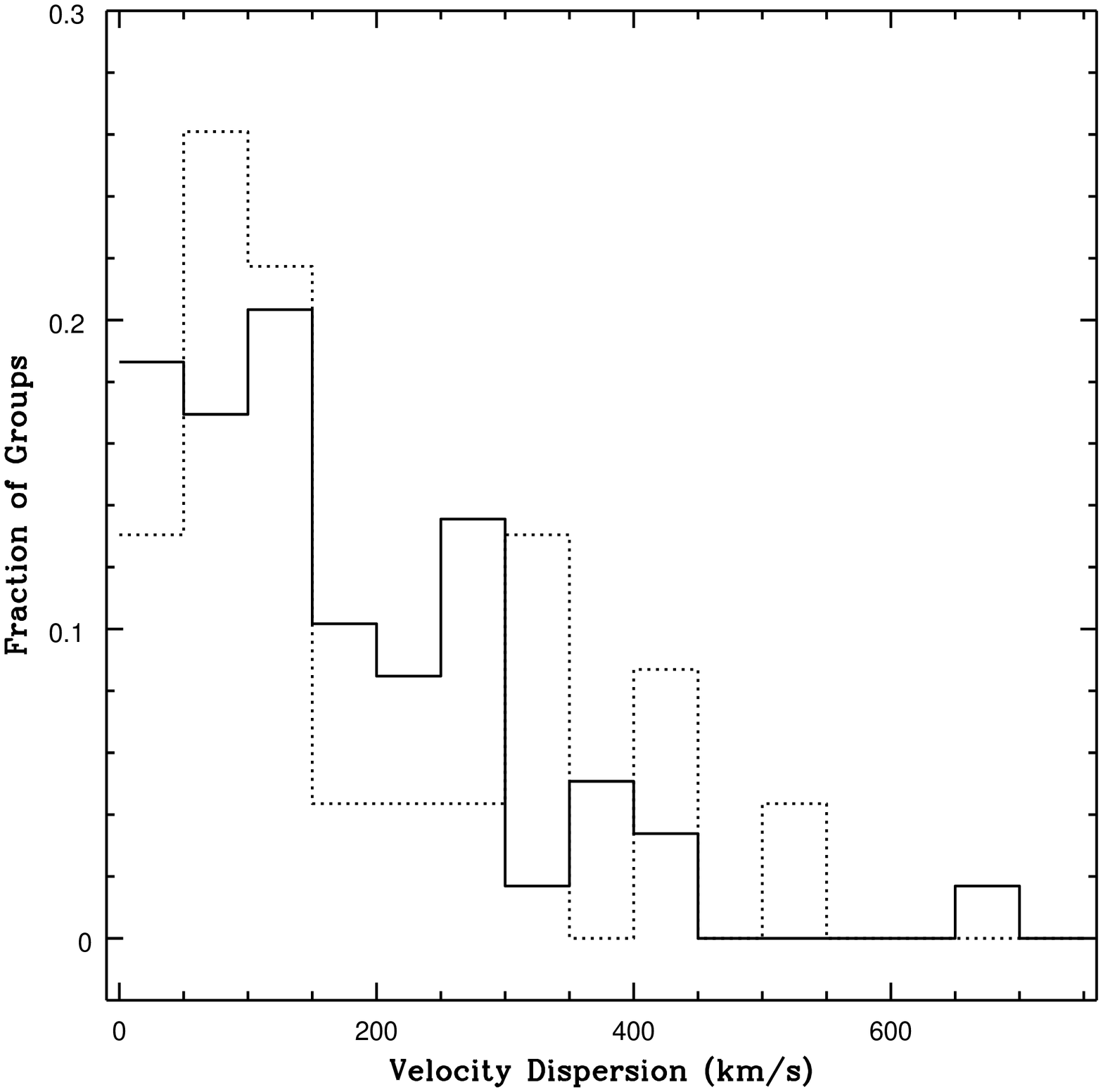}{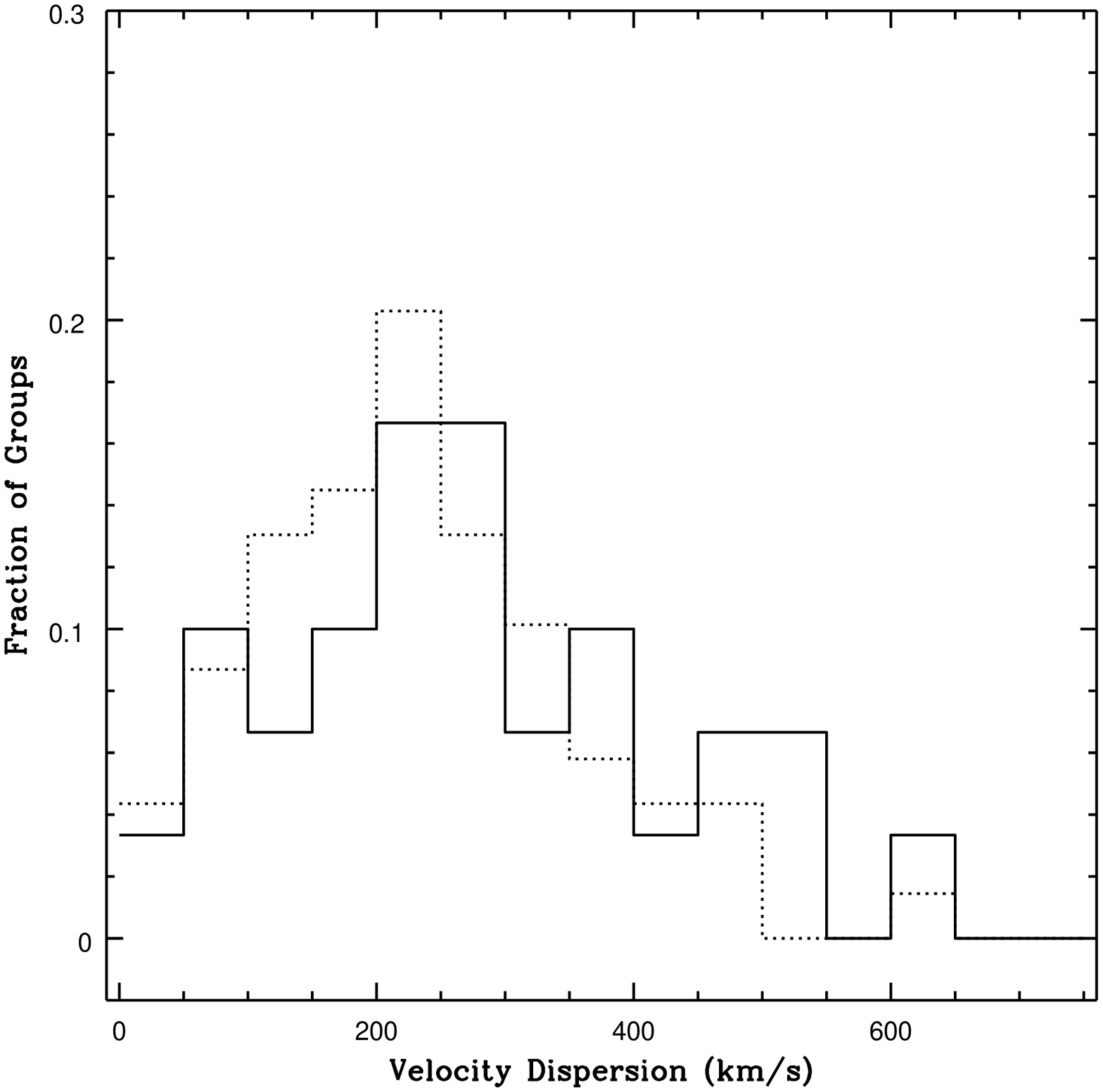}
\caption{Velocity dispersion distributions of compact groups: (a) RSCG's with $N=3$ (solid line) and all 
HCG's with $N=3$ (dotted line), and (b) all RSCG's with $N \geq 4$ (solid line) and all 
HCG's with $N \geq 4$ (dotted line).} \label{fig:veldisp}
\vspace{12pt}
\plottwo{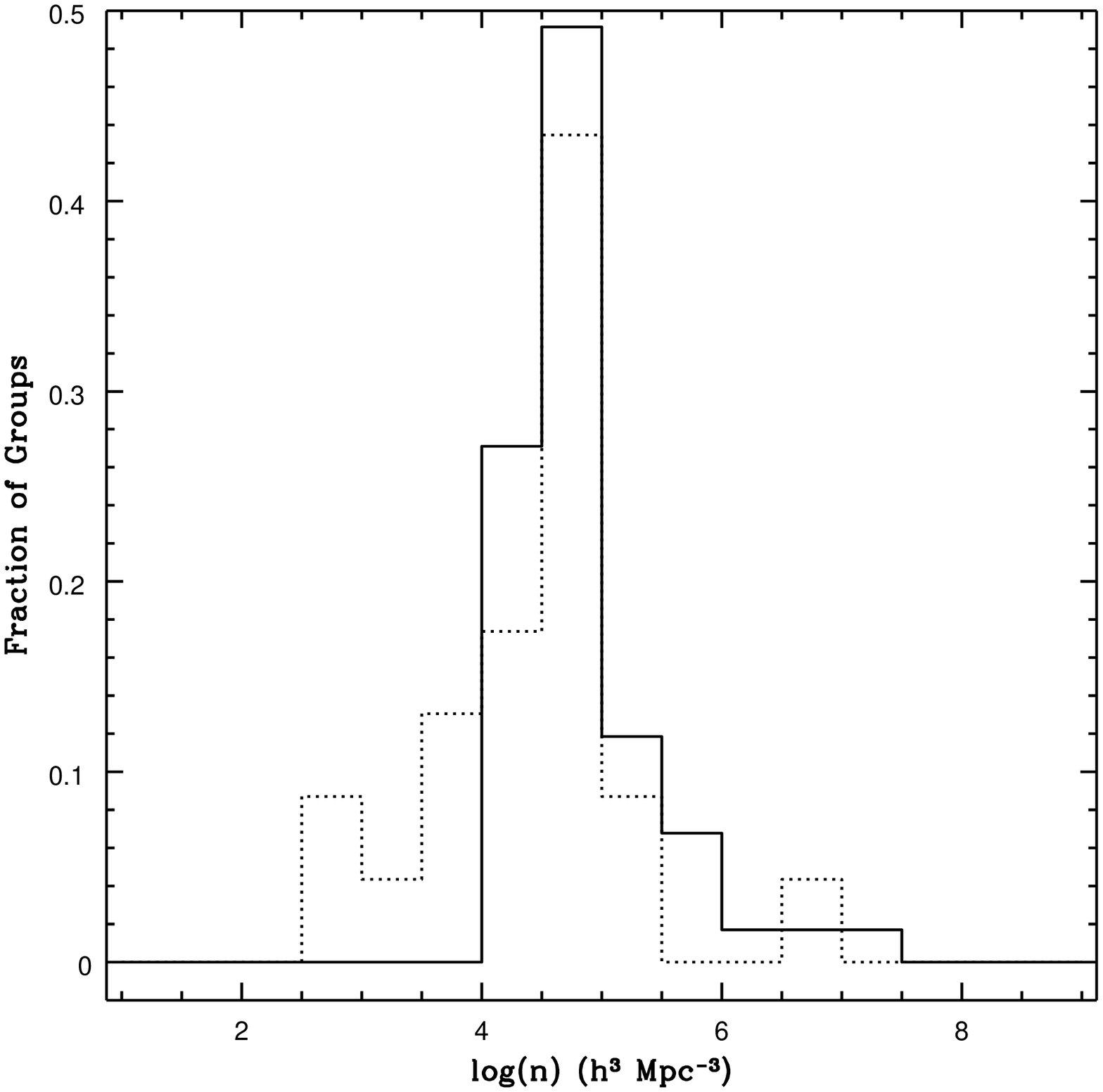}{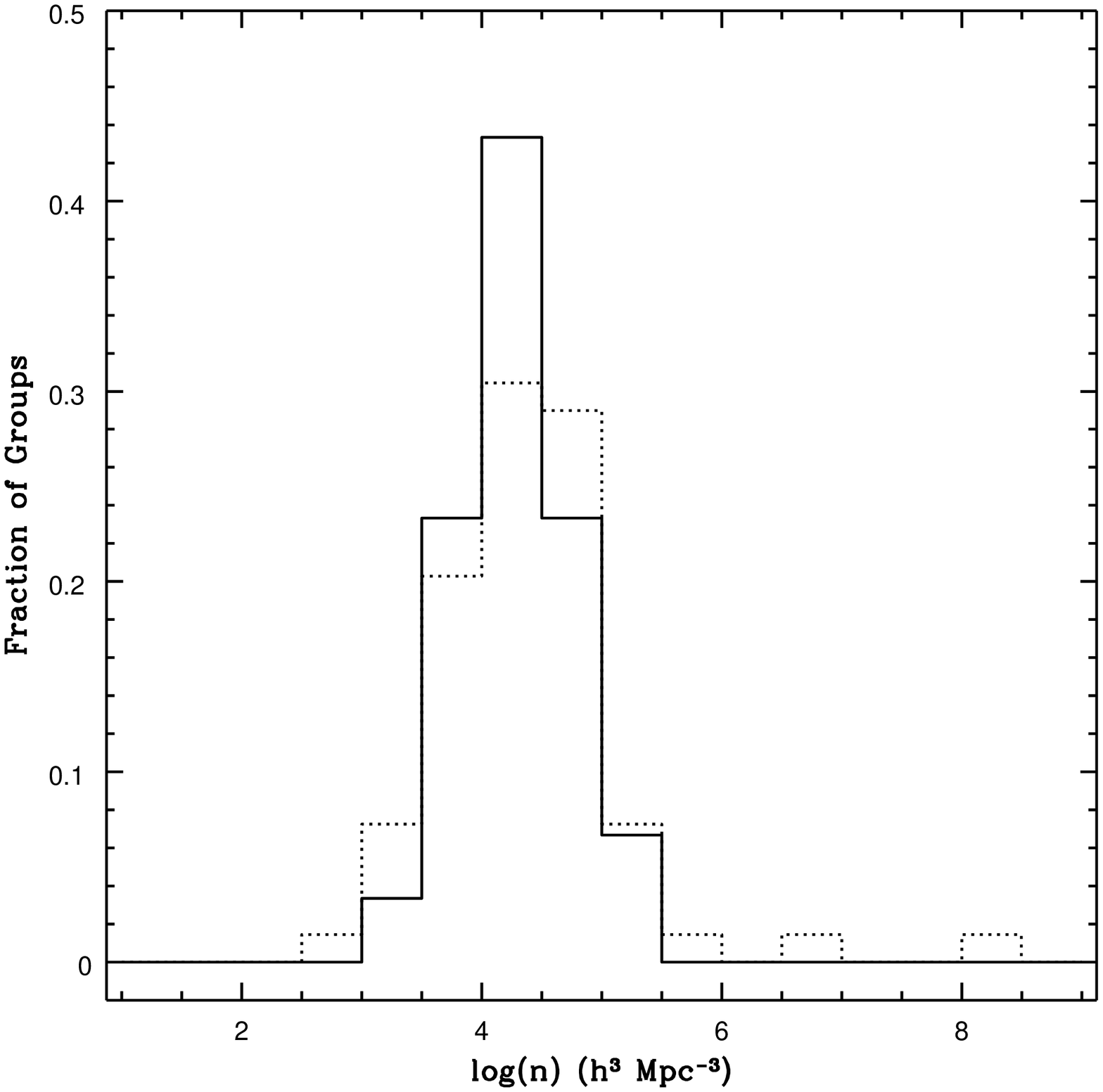}
\caption{Density distributions of compact groups: (a) all 
RSCG's with $N=3$ (solid line) and all HCG's with $N=3$
(dotted line), and (b) all RSCG's with $N \geq 4$
(solid line) and all HCG's with $N \geq 4$ (dotted line).} \label{fig:dens}
\end{figure}

We also compare the group densities.  We choose the
density statistic $n=\frac{3N}{4 \pi R^{3}}$, with $N$ 
the number of detected galaxies and
$R$ the radius of the smallest circle which encloses all the
galaxy centers in the RSCG.  
Figures~\ref{fig:dens}a and~\ref{fig:dens}b show
the distribution of $\log(n)$ for both the HCG's (solid line) and
all RSCG's (dotted line).  
Figure~\ref{fig:dens}a includes only groups with $N= 3$ and Figure~\ref{fig:dens}b 
includes only
groups with $N \geq 4$.
Table~\ref{tab:dens} lists the median, first and third quartiles 
of these distributions.  
As expected, the RSCG density distributions
agree well with the HCG density distributions at high densities.

The density distributions of 
our compact groups fall off sharply below $\log(n) \approx 4$,
where $n$ is in Mpc$^{-3}$, especially
for groups with $N = 3$.  
This cutoff is an artifact of our search algorithm.  
Any galaxy in a compact group has a projected separation of at most
$50\ {\rm kpc} = 0.05\ {\rm Mpc}$ from at least one other 
galaxy in its group.  For a group of 3, the minimum 
density allowed is then 3 galaxies in a line, 
separated by 0.05~Mpc; the
group radius is $R=0.05$ Mpc.  The minimum group density is
then $n= 5.7 \times 10^{3}$ Mpc$^{-3}$; $\log(n_{\rm min}) = 3.76$.
A similar calculation for a group
of 4 galaxies yields $\log(n_{\rm min}) = 3.35$.  
In general, the search algorithm introduces a complex selection as a 
function of density, but this selection is only relevant over a small
range of densities.
The HCG's do not display this
sharp cutoff because selection is based on angular rather than 
physical extent.
These criteria lead to a range of HCG densities corresponding to a 
range of distances.

\section{Hickson's Criteria}

Unlike Hickson, we do not include isolation and luminosity criteria in
our selection algorithm.  Here we investigate the physical implications of
the application of Hickson's more restrictive criteria.  

\subsection{Population and Surface Brightness Criteria}

Hickson limited group membership to galaxies with $m \leq m_{1}+3$
because galaxies with similar magnitudes are more likely 
to be at the same redshift than galaxies with a broader magnitude 
distribution.  This problem is irrelevant for groups selected in
redshift space.  Furthermore, faint galaxies are probably less
massive and hence less important dynamically.  This effect is relevant to 
very nearby groups in our sample.  Seven of our groups
(all with median velocities
$\leq  1600\ {\rm km \ s}^{-1}$) contain fewer than three
members in the interval [$m_{1}$,$m_{1} +3$].  Most of these groups
are located in the outskirts of the Virgo cluster [see, for example,
Mamon \markcite{m89} (1989), who finds RSCG 65 in Virgo].  In
general, these nearby systems contain one or two large galaxies 
along with small
satellites; they are therefore different in character from 
the other RSCG's and from the HCG's.  To exclude these systems, in later Sections we ignore nearby
groups ($cz \leq 2300$~km~s$^{-1}$) when exploring the environments
of RSCG's and when calculating
the compact group selection function.

Hickson required $\overline{\mu}_{G} < 26.0$, where $\overline{\mu}_{G}$
is the surface brightness in E, for the POSS plates,  of the group in
magnitudes arcsec$^{-2}$.  
We translate this criterion as  $\overline{\mu}_{G,{\rm Zw}} < 27.7$ 
using arguments from 
Prandoni, Iovino \& MacGillivray \markcite{p94} (1994) that 
a surface brightness of
26.0 magnitudes arcsec$^{-1}$ in $E$ is equivalent 
to a surface brightness of 27.7 magnitudes arcsec$^{-1}$ in $b_{j}$ which
is similar to the Zwicky magnitude scale.
Four of our compact groups do not meet this requirement.  
All of these groups are in the CfAnorth survey.
They are nearby ($cz \leq$ 1200~km~s$^{-1}$), and have large angular radii
($\theta_{G} > 14 '$).  All of
these groups fail the isolation criterion as well.

\subsection{Isolation Criterion}

Hickson's isolation criterion has complex physical
implications.  In a survey of compact groups selected on the sky, 
interloping galaxies can appear in the isolation annulus 
($\theta_{G} < \theta < 3\theta_{G}$) and thus can cause 
bonafide systems to fail
the isolation criterion.  

Because we have a complete catalog in redshift
space, we can apply a cleaner isolation criterion in three dimensions.
We compute the isolation ratio,
$\Upsilon \equiv \theta_{N}/\theta_{G}$, 
for all groups in the sample. 
$\theta_{G}$ is again the radius of the smallest circle which contains
the centers of all the group members and $\theta_{N}$ is the angular radius
of the largest concentric circle which contains no further galaxies
within 1200~km~s$^{-1}$ of the median group velocities.  Our velocity 
separation cutoff is somewhat arbitrary; it is more relaxed than the group
selection cutoff to allow for a larger velocity dispersion in the group
environment.
Our isolation
criterion is not completely
equivalent to Hickson's criterion because we only search for surrounding
galaxies brighter than the magnitude
limit, $m_{\rm Zw}=15.5$.  Thus, for groups with 
$m_{1} > 12.5$, we do not search for surrounding galaxies
throughout $[m_{1},m_{1}+3]$.
Twenty-five RSCG's fail the ``isolation criterion'' which requires
$\Upsilon > 3$.  Six of these groups have  $N=3$, 19 have $N \geq 4$.

\begin{figure}[p]
\centerline{\epsfxsize=3.7in%
\epsffile{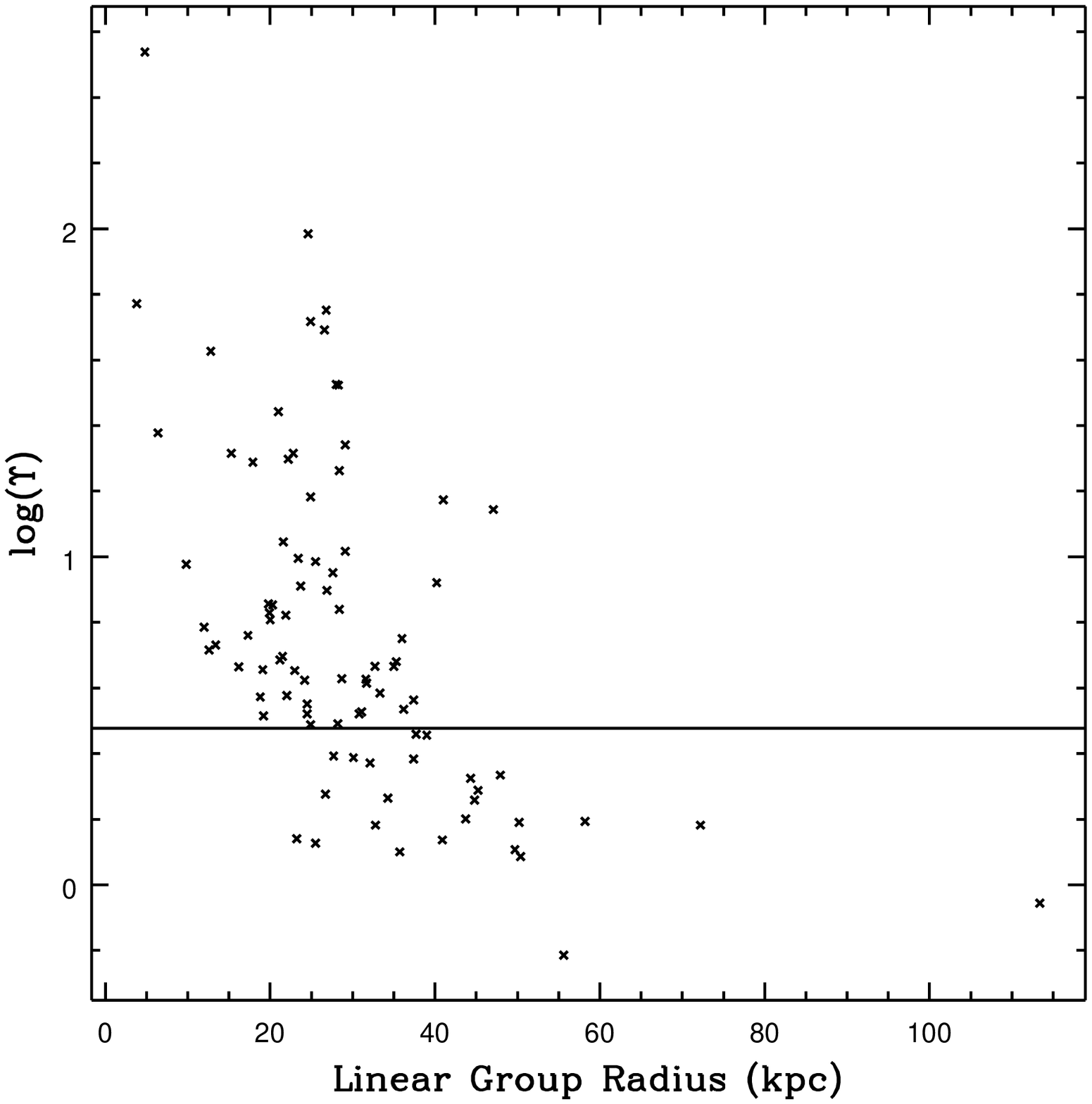}}
\caption{Dependence of isolation  on group radius:
log($\Upsilon$) vs. radius in kpc.
The horizontal line is the boundary between groups
that satisfy the ``isolation criterion'' and groups that do not. } \label{fig:isovrad}
\vspace{12pt}
\plottwo{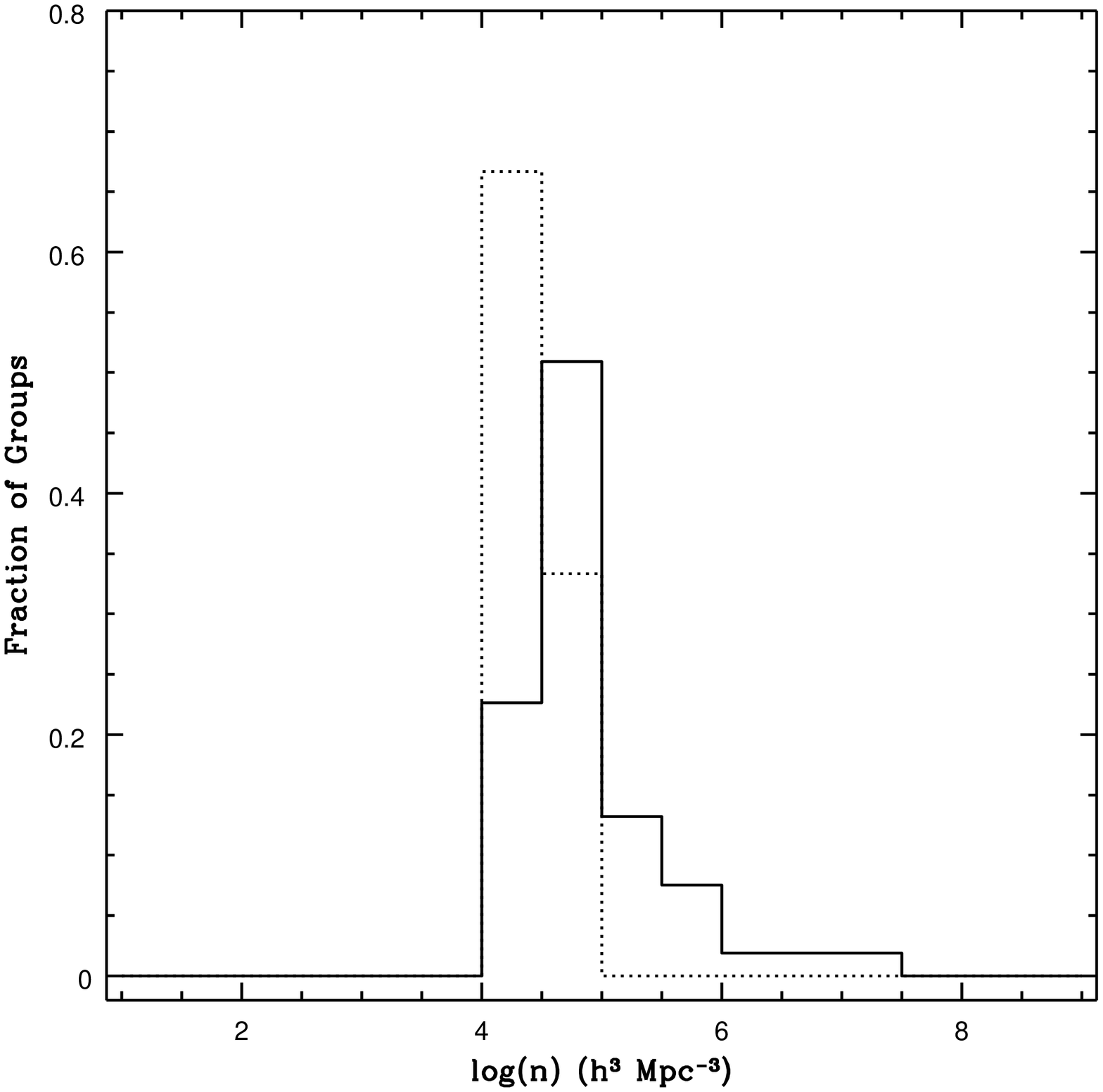}{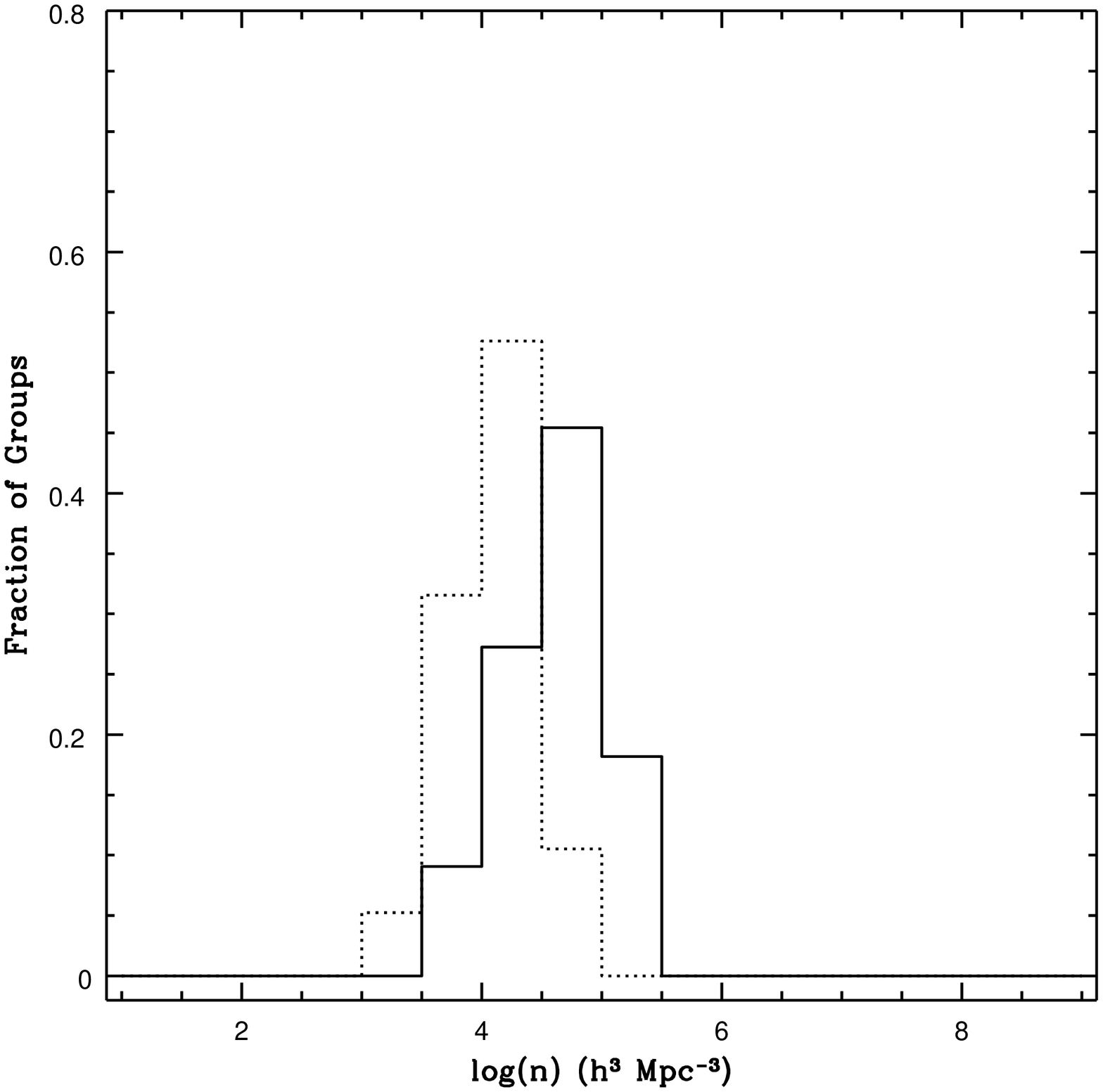}
\caption{Density distributions of RSCG's: (a) all isolated 
compact groups with $N=3$ (solid line)  and non-isolated 
compact groups with $N=3$ (dotted line), and (b) all isolated compact groups with 
$N \geq 4$ (solid line) and non-isolated compact groups with 
$N \geq 4$ (dotted line). } \label{fig:isodens}
\end{figure}

We  explore the differences between ``isolated'' 
groups and groups with $\Upsilon < 3$.
The isolation parameter $\Upsilon$ 
is a {\it strong} function of group radius.  
Figure~\ref{fig:isovrad} shows $\log(\Upsilon)$ as a
function of projected group radius in kpc.  
The horizontal line is $\Upsilon = 3$.
Groups with large angular radii are
not isolated simply because their isolation annuli are larger.  
Because of this dependence, 
$\Upsilon$ alone does not specify the physical
environment of a particular RSCG.
In fact, there is evidence that most HCG's are also embedded in larger
systems (de Carvalho et al.\ \markcite{c94} 1994; Ramella et al.\ 
\markcite{r94} 1994).

The correlation of $\Upsilon$ with radius 
results in a correlation of $\Upsilon$ with group density.  
Figures~\ref{fig:isodens}a and~\ref{fig:isodens}b 
show the density distributions of the
two subsamples (isolated and not isolated)
for groups with $N=3$ and with $N \geq 4$, respectively.
We apply a K-S test to the
density distributions of the two subsamples in each case. 
The null hypothesis that the density values are drawn
from the same distribution is acceptable only at the  3.6\% significance level
for groups with $N=3$ and the  0.46\% level for groups with  $N \geq 4$.
Thus isolation does imply higher density, as expected.  
The correlation is not physical; it is a result of the correlation of 
$\Upsilon$ with group radius.
 
\begin{figure}[t]
\vspace{72pt}
\plottwo{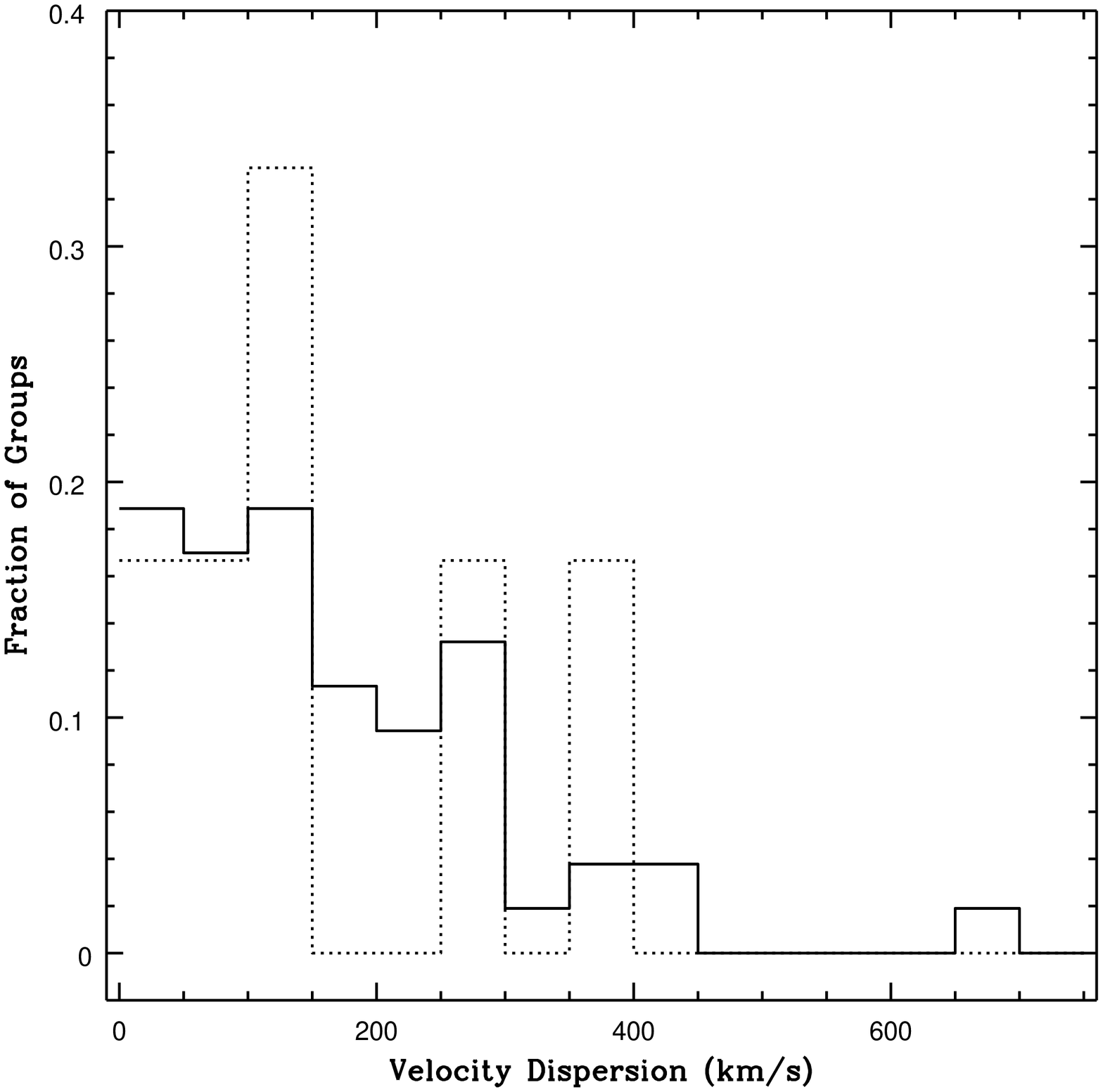}{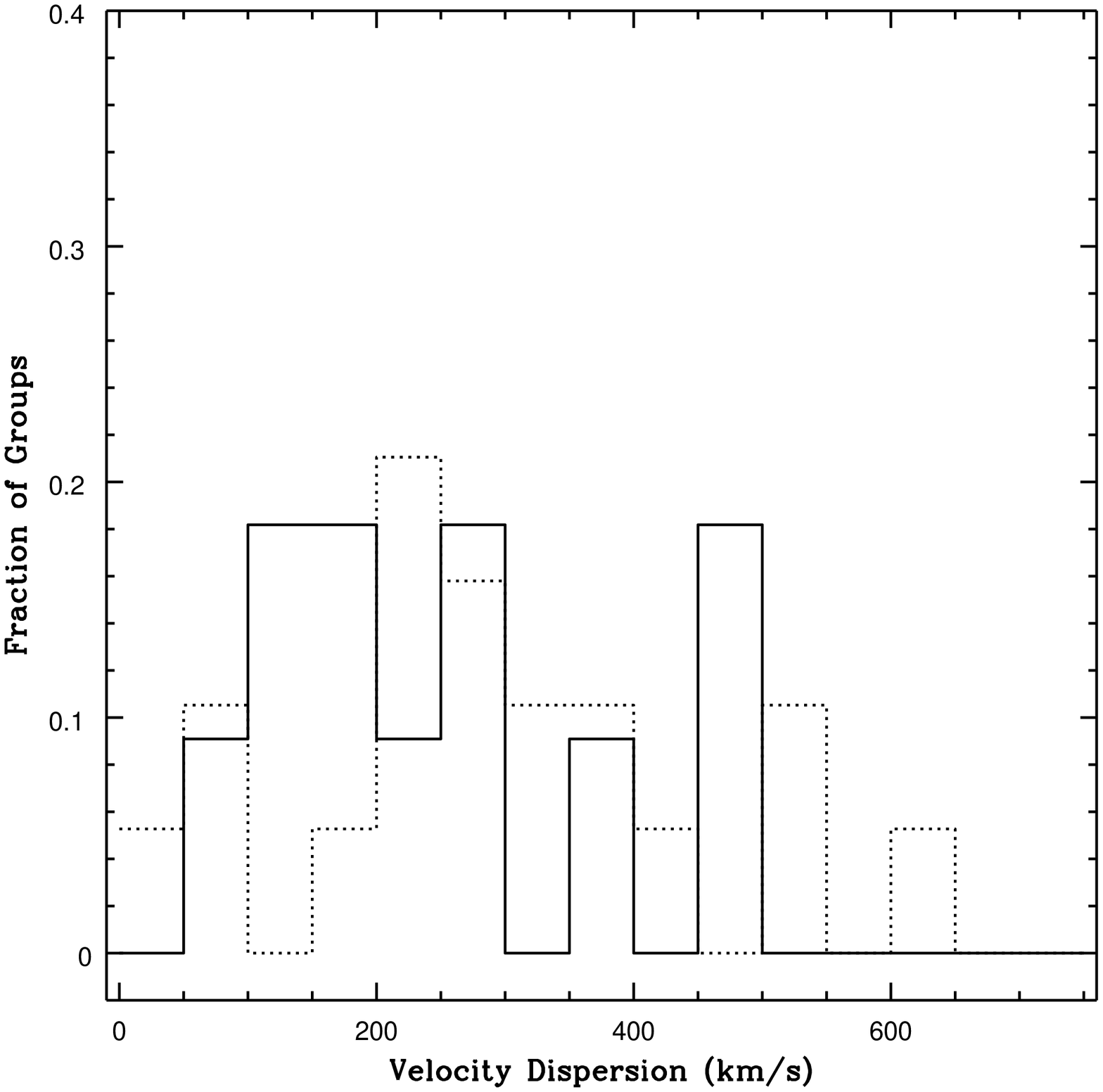}
\caption{Velocity dispersion distributions of compact groups: (a) all isolated 
RSCG's with $N = 3$ (solid line) and non-isolated 
RSCG's with $N=3$ (dotted line), and
(b) all isolated RSCG's with $N \geq 4$
(solid line) and non-isolated RSCG's with $N \geq 4$ (dotted line).} \label{fig:isovel}
\end{figure}

The velocity distributions are consistent for the two 
samples.  The velocity
dispersion distributions of the subsamples agree at the   99\% 
confidence level for $N=3$ and  at the 49\% confidence level for $N \geq 4$.  
Figures~\ref{fig:isovel}a and~\ref{fig:isovel}b 
show the velocity distributions for
groups with $N=3$ and groups with $N \geq 4$, respectively.
We note, however, that the subsamples of RSCG's in these comparisons are
small.   
There are 11 groups with $N \geq 4$ and $\Upsilon > 3$, 
19 groups with $N \geq 4$ and $\Upsilon < 3$, 
53 groups with $N=3$ and $\Upsilon > 3$, 
and only 6 groups with $N=3$ and $\Upsilon < 3$. 
 
We visually examine regions around
the 48 RSCG's with $\Upsilon > 3$ and $cz > 2300\ {\rm km\ s}^{-1}$
using Digitized Sky Survey images.  About 1/3 of these
groups have faint galaxies (with $m \lesssim m_{1}+3$)
in their isolation annuli without measured redshifts.
In other words, at least 30 of the RSCG's are isolated according to
Hickson's criteria.  However, this apparent isolation is not a good 
predictor of the environment of compact systems and contains no
information about the physics of the systems.

\section{Environments of Compact Groups}

Previous studies of the environments of compact groups of
galaxies yield mixed messages; some investigators 
find that compact groups are in
dense regions and others conclude that they are not
(Rood \& Williams \markcite{ro89} 1989; Rubin, Hunter \& 
Ford \markcite{r91} 1991; de Carvalho, Ribeiro \& Zepf 
\markcite{c94} 1994; Ramella et al.\ \markcite{r94} 1994).
These discrepant conclusions are at least in part a result of incomplete
data sets.
We investigate the environments of the RSCG's in redshift space. We find that
most of them are embedded in loose groups or clusters.

Using the method
of Ramella et al.\ \markcite{r94} (1994), 
we count the number of galaxies (including 
the galaxies in the RSCG), 
$N_{n}$, in the redshift survey within 
1.5~$h^{-1}$ Mpc (projected) of the group
center and within 1500~km~s$^{-1}$ of the median group velocity, $v_{\rm med}$.
These scales are well-matched 
to the size of loose groups (Ramella, Geller \&
Huchra \markcite{r89} 1989).

For the remaining analyses, we exclude the 31 RSCG's with 
redshifts less than 2300~km~s$^{-1}$.  Many of these systems differ
significantly from more distant RSCG's, making them difficult to model.
They are often 
composed of a bright galaxy and its very faint satellites (see
\S 6.1), a combination that we cannot detect at 
higher redshift.  In addition, these groups are hard to model because they are
within the local supercluster.

Table~\ref{tab:env} lists $\frac{N_{n}}{N_{\rm int,env}}$ for RSCG's with 
average redshifts greater than 2300~km~s$^{-1}$.  
$N_{\rm int,env}$ is the expected
number of interlopers within 1.5 $h^{-1}$ Mpc of the group center  
in the velocity 
interval $[v_{\rm med}-1500,v_{\rm med}+1500]$;  
\begin{displaymath}
N_{\rm int,env} \approx \int_{V_{\rm env}} \overline{\rho} dV,
\end{displaymath}
where $\overline{\rho}$ is the 
average density of the relevant portion of the survey
and $V_{\rm env}$ is the volume of
the environment of the group. 
Similarly, 
\begin{displaymath}
N_{n} \approx \int_{V_{\rm env}} \rho_{\rm env} dV, 
\end{displaymath}
where $\rho_{\rm env}$ 
is the density of the environment of the group. 
Therefore,
\begin{displaymath}
\frac{N_{n}}{N_{\rm int,env}} \approx \frac{ \langle\rho_{\rm env}\rangle}
{\langle \overline{\rho} \rangle},
\end{displaymath}
where
the brackets denote averaging over the volume of the
environment of the RSCG.
To calculate this expected number of interlopers we integrate over 
the {\it true} luminosity
function (Table~\ref{tab:lumrd}).  
We use the true luminosity function instead of the observed one for ease
of calculation; the error introduced is small.
Therefore,
\begin{equation}
N_{\rm int,env}=\int_{v_{\rm med}-1500}^{v_{\rm med}+1500}  
\int_{-\infty}^{M_{\rm lim}(v)} \pi \left( \frac{1.5\  {\rm Mpc} \times v}{v_{\rm med}}\right)^{2}
\Phi_{\rm true}(M) dM \frac{dv}{H_{0}}, \label{eqn:nint}
\end{equation}
where $M_{\rm lim}(v)$ is the limiting absolute magnitude at redshift
$v$, $\Phi_{\rm true}$ is the true luminosity function of the
relevant survey and $v_{\rm med}$ is the median RSCG velocity.
When calculating $N_{\rm int, env}$ we integrate over 1.5 $h^{-1}$ projected
Mpc, although some groups are centered on the edge of the
survey where we do not observe all galaxies in the neighborhood.
We indicate these groups in the table.

We also list $N_{\rm int,cg}$ in Table~\ref{tab:env}.  We calculate this number
by replacing 1.5~Mpc with $R_{\rm cg}$, the group radius in Mpc, in 
Equation~\ref{eqn:nint}.  We then compute the overdensity of the 
RSCG, 
$\frac{N_{\rm cg}}{N_{\rm int,cg}} \approx \frac{\langle \rho_{\rm cg} \rangle}
{\langle \overline{\rho} \rangle }$. 
$N_{\rm cg}$ is the number of galaxies in the RSCG and $\rho_{\rm cg}$ is the density 
of the compact group.  The RSCG's 
have $460 < \frac{\langle \rho_{\rm cg} \rangle}
{\overline{\langle \rho}\rangle } < 150,000$.

For  $cz \geq 2300$~km~s$^{-1}$, 43  RSCG's (74\%) 
have $\langle \rho_{\rm env}  \rangle \ \gg \ \langle \overline{\rho} \rangle$.
This result gives the same impression as the conclusion of
Ramella et al.\ that
most but not all HCG's are embedded in
looser systems.  

All 10 RSCG's with $cz \geq 2300\ {\rm km\ s}^{-1}$
which fail the ``isolation'' criterion 
(\S 6) satisfy $N_{n} > N_{\rm int} + 3\sqrt{N_{\rm int}}$, as expected.
However, 30 {\it isolated} RSCG's with $cz \geq 2300\ {\rm km\ s}^{-1}$ 
also satisfy $N_{n} > N_{\rm int} + 3\sqrt{N_{\rm int}}$ 
(for $m_{\rm Zw} \leq$~15.5) --- another indication that
``isolation'' is not physically meaningful.

\begin{figure}
\plotfiddle{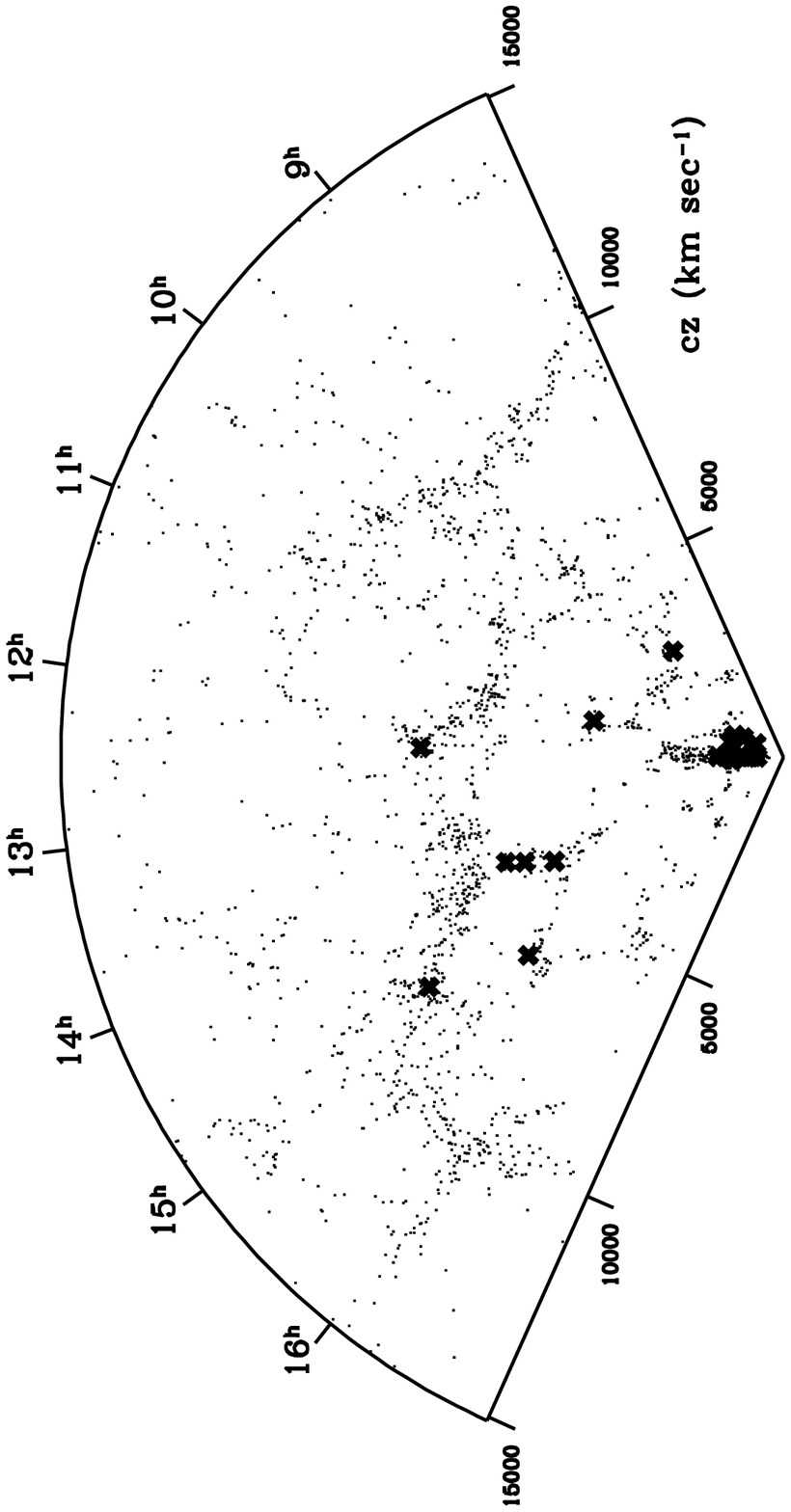}{6cm}{-90}{70}{70}{-275}{400}
\caption{Distributions of redshift survey
galaxies ($\cdot$) and RSCG's ($\times$). CfAnorth survey: 
$8.5 ^\circ < \delta < 17.5 ^\circ$. } \label{fig:no1}
\end{figure}

\begin{figure}
\plotfiddle{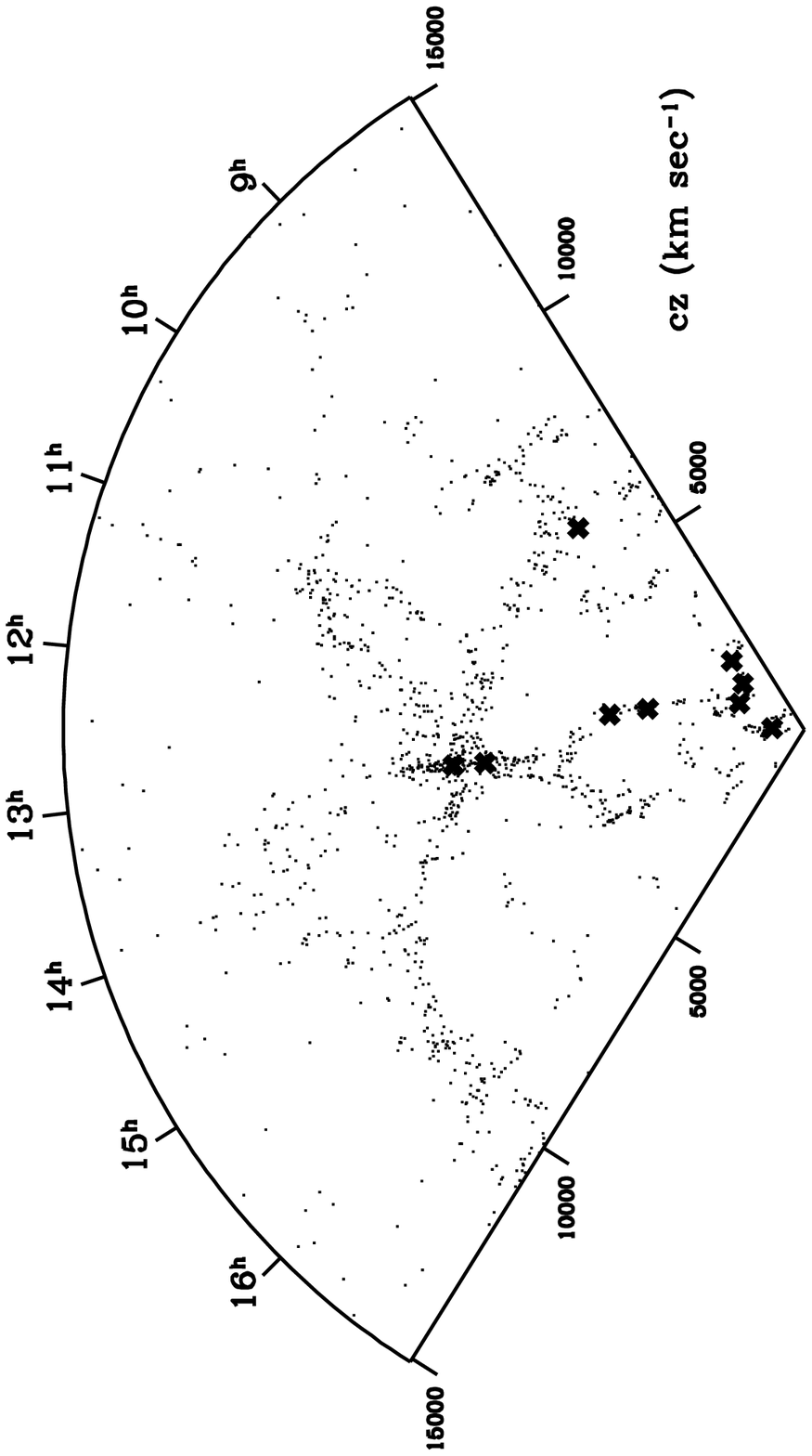}{6cm}{-90}{70}{70}{-275}{400}
\caption{Distributions of redshift survey
galaxies ($\cdot$) and RSCG's ($\times$).
CfAnorth survey: $26.5 ^\circ < \delta < 35.5 ^\circ$. } \label{fig:no3}
\end{figure}

\begin{figure}
\plotfiddle{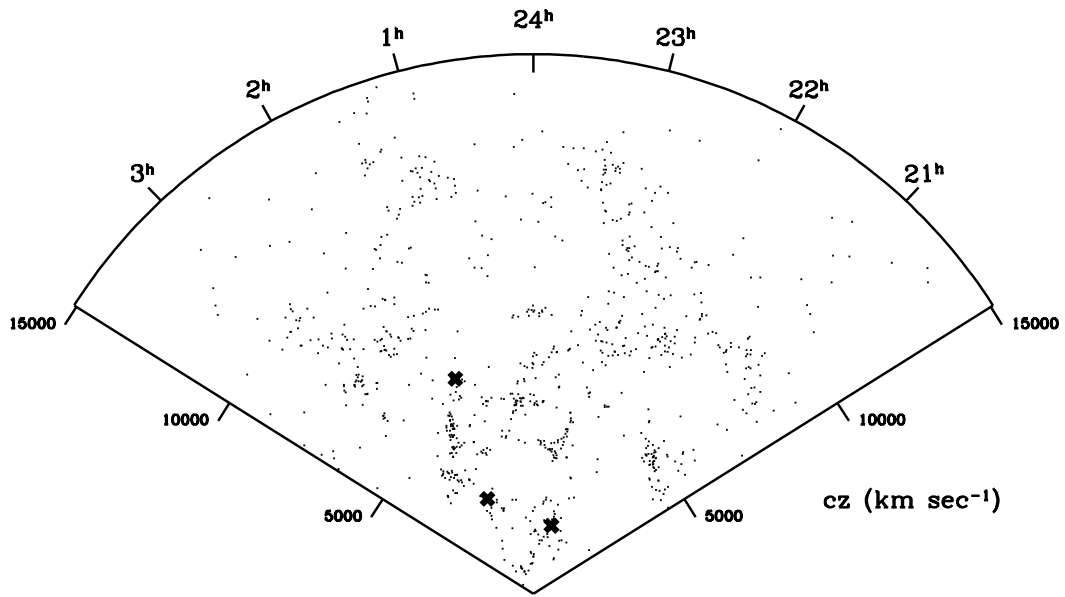}{6cm}{-90}{70}{70}{-275}{400}
\caption{Distributions of redshift survey
galaxies ($\cdot$) and RSCG's ($\times$).
CfAsouth survey: $10 ^\circ < \delta < 22 ^\circ$. } \label{fig:so2}
\end{figure}

\begin{figure}
\plotfiddle{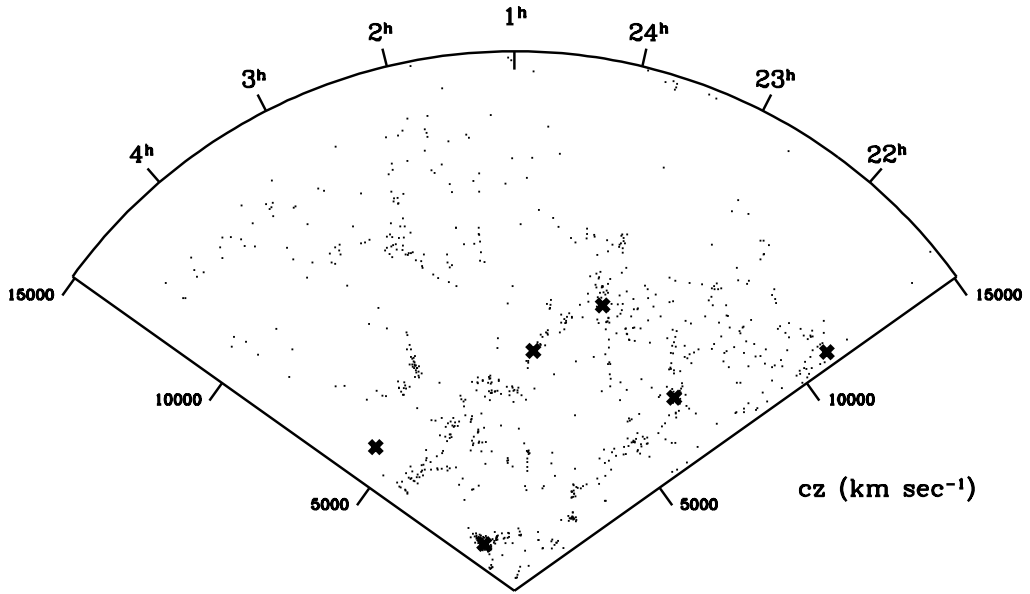}{6cm}{-90}{70}{70}{-275}{400}
\caption{Distributions of redshift survey
galaxies ($\cdot$) and RSCG's ($\times$).
SSRS2 survey: $ -30 ^\circ < \delta < -20 ^\circ$. } \label{fig:ss2}
\end{figure}

Figures~\ref{fig:no1}~--~\ref{fig:ss2} show examples of the embedding of
RSCG's within the large-scale structure of the nearby universe.
They show galaxies with redshifts
in the interval (300~km~s$^{-1}$,~15,000~km~s$^{-1}$)
in the redshift surveys ($\cdot$) and RSCG
centers ($\times$) (including RSCG's with
$cz < $~2300~km~s$^{-1}$).  Compact groups generally follow
the large-scale distribution of galaxies, often appearing in the
densest regions of the survey.  However, Figure~\ref{fig:ss2} shows
the seemingly rare example of a compact group in an apparent void.

Most RSCG's are
embedded in larger systems regardless of how ``isolated'' they 
appear.  
The isolation criterion is unphysical:  groups with
smaller radii are more likely to appear ``isolated'' mainly because their 
isolation annuli are smaller.  This variation with radius leads to
a dependence of density, $n$, on isolation only because density is computed
using the group radius; thus the correlation of $\Upsilon$ with
$n$ is unphysical. From here on we include all 
RSCG's  whether they satisfy the isolation criterion or not.

\section{Abundance of Compact Groups}

The abundance of compact groups has important physical 
implications for cosmology.  The compact group abundance can indicate whether 
compact groups are bound systems or chance 
alignments of galaxies within loose groups or along filaments.
If CG's are truly compact systems, the abundance of these
systems is dependent on the density parameter, $\Omega_0$.  If $\Omega_0 \sim 1$
these systems should be forming in the present epoch. Furthermore,
the abundance of compact systems should be related to the merger rate
at the current epoch.

We calculate the volume number density of RSCG's.
We compute abundances using all RSCG's and using only 
RSCG's with $N \geq 4$.    We consider only systems
with median redshifts $>$~2300~km~s$^{-1}$.  

\subsection{The Luminosity Distribution of Galaxies in Compact 
Groups}

\begin{figure}
\plotone{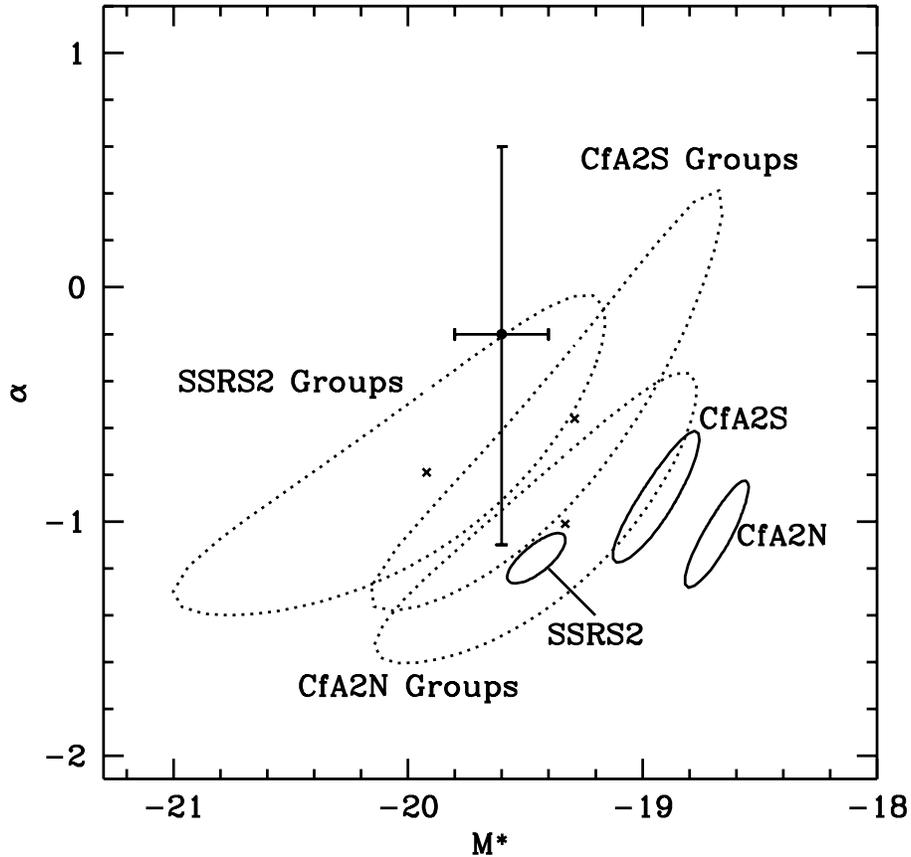}
\caption{Confidence levels (ellipses) for Schechter function parameters describing
separately 
the SSRS2, CfAnorth and CfAsouth RSCG's (dotted line, $1 \sigma$)
and the confidence levels (solid line, $2 \sigma$) for the entire CfAnorth and CfAsouth  surveys 
and the SSRS2 survey (Marzke, Huchra \& Geller 1994;
da Costa et al.\  1994).  
Mendes de
Oliveira \& Hickson (1991) HCG LF parameters are shown as a single
point with errors.} \label{fig:lum1}
\end{figure}

To model
the selection function of compact groups, we have to evaluate the luminosity
function (LF) for compact group members.
We compute
the Schechter \markcite{s76} (1976) function parameters 
(Marzke, Huchra \& Geller \markcite{m94} 1994, 
Efstathiou, Ellis \& Peterson \markcite{e88} 1988, Loveday et al.\ 
\markcite{l92} 1992),
$\alpha_{\rm cg}$ and $M_{\star, {\rm cg}}$, for the
CfAnorth, CfAsouth and SSRS2 RSCG samples.  Table~\ref{tab:lum} lists
$\alpha_{\rm cg}$ and $M_{\star,{\rm cg}}$ for the CfAnorth, CfAsouth and SSRS2 catalogs of
compact group galaxies.  The samples include 84 galaxies, 77 galaxies and
34 galaxies, respectively.
In all cases, $M_{\star,{\rm cg}} < M_{\star,{\rm survey}}$
and $\alpha_{\rm survey} < \alpha_{\rm cg}$, where $M_{\star,{\rm survey}}$ and 
$\alpha_{\rm survey}$ are the Schechter function parameters for each 
region of the complete redshift survey (Table~\ref{tab:lumrd}).  
In other words, the characteristic luminosity is brighter
and the faint end shallower for the compact groups.
Figure~\ref{fig:lum1} illustrates the significance of these differences; it
shows 1$\sigma$ contours for the compact group galaxies (dotted lines)
and 2$\sigma$ contours for all CfAnorth, CfAsouth,
and SSRS2 survey galaxies (solid lines).  
These differences are similar for all portions of the redshift survey.
The 1$\sigma$ error ellipse for the compact group galaxies does
not overlap the 2$\sigma$ survey error ellipse
for any of the surveys.

Mendes de Oliveira \& Hickson \markcite{m91} (1991), Prandoni, Iovino \&
MacGillivray \markcite{p94} (1994), and Ribeiro, de Carvalho \& Zepf 
\markcite {rr94} (1994)
have also compared 
compact group LF's with the ``field''. 
Mendes De Oliveira \& Hickson \markcite{m91} (1991) used simulations of 
compact groups to derive an HCG galaxy
LF.  They fit a Schechter function with
parameters $\alpha = -0.2^{+0.8}_{-0.9}$ and 
$L_{\star}=1.1 \pm 0.2 \times 10^{10} h^{-2} L_{\sun}$ 
($M_{\star,B} =-19.6 \pm 0.2$).  We show the value of Mendes de
Oliveira \& Hickson,
with error bars, in Figure~\ref{fig:lum1}. The error bars 
of Mendes de Oliveira \& Hickson overlap 
the 1$\sigma$ LF parameter error ellipses of all three RSCG subsamples.  
Mendes De Oliveira \& Hickson claim a 
significant deficiency of low-luminosity galaxies, as compared to field, loose
group and cluster galaxies.  Prandoni, Iovino \& MacGillivray \markcite{p94} 
(1994) and Ribeiro, de Carvalho \& Zepf \markcite {rr94} do not confirm this deficiency, although their samples may be contaminated by superimposed background galaxies.  
Ribeiro, de Carvalho \& Zepf \markcite {rr94} 
compare faint galaxy counts
inside and outside 22 HCG's. 
They find that the LF inside HCG's is consistent with the field LF.

\subsection{The Selection Function of Compact Groups}

For the following analysis, we 
use the survey LF's when estimating the
compact group selection function because the survey LF's are much better
determined and better understood.  
We thus underestimate the selection function of compact groups somewhat and we
overestimate the compact group density slightly, assuming that the differences
in luminosity functions are real.

We require that each compact group contain $j$ or more galaxies 
coincident in redshift space, where $j=3$ is our criterion for
selection; we also restrict our sample for comparison to
Hickson's (1982) by adopting $j=4$.
We thus estimate the selection function for RSCG's analytically,
by modeling the probability of detecting the $j$th brightest galaxy.
We assume that the galaxies in RSCG's are drawn
at random from a 
magnitude distribution of fixed form: $\overline{\Phi}(M)$. 
We calculate
$P_{j}(M)dM$, the probability that the $j$th brightest member of
a group of galaxies lies in the interval [$M$, $M+dM$]; 
this probability is proportional to
$P_{(j-1)<M}(M) \overline{\Phi}(M) dM$, 
where $P_{(j-1)<M}(M)$ is the probability that exactly
$j-1$ members of the group are brighter than $M$.  If $\lambda_{M}$
is the average number of galaxies in a group brighter than $M$,
then,
\begin{equation}
\lambda_{M}=\kappa \int_{-\infty}^{M} 
\overline{\Phi}(M')dM', \label{eqn:kappa}
\end{equation}
where, in general, $\kappa = \kappa(M)$ is a normalization parameter.
Then from the Poisson distribution we derive
\begin{equation}
P_{(j-1)<M}=\frac{e^{-\lambda_{M}}\lambda_{M}^{(j-1)}}{(j-1)!}.
\label{eqn:2gtm}
\end{equation}
For a survey with a limiting apparent magnitude of 15.5, the 
limiting absolute magnitude at any redshift
$v=cz$ is $M_{\rm lim}(v)=-9.5-5\log_{10}(\frac{v}{H_{0}})$, where
the dimensions of $\frac{v}{H_{0}}$ are Mpc.  
Therefore, the probability
of detecting $j$ or more members of a group of galaxies is
\begin{equation}
P_{\rm detection}(v)=\frac{1}{\aleph} \int_{-\infty}^{M_{\rm lim}(v)} P_{j}(M)dM=
\frac{1}{\aleph}
\int_{-\infty}^{M_{\rm lim}(v)} \frac{e^{-\lambda_{M}}\lambda_{M}^{(j-1)}}{(j-1)!} 
\overline{\Phi}(M) dM.
\label{eqn:pdet}
\end{equation}
The factor $\aleph$ is the normalization of $P_{\rm detection}(v)$ which ensures
that $P_{\rm detection}(0)=1$.

\subsubsection{Estimating the $\kappa$ parameter}

In order to evaluate the probability of detecting a compact group
[Equation~\ref{eqn:pdet}] we assume that $\kappa(M)$ in 
Equation~\ref{eqn:kappa} is a constant.  In other words, we assume that the
number of galaxies brighter than $M$ in a compact group, $\lambda_M$,
is proportional to the integral of the luminosity distribution of galaxies and
that the normalization ($\kappa$) is the same for every compact group.
These assumptions are not strictly correct.  
The inaccuracy introduced
by approximating $\lambda_M$ by a smooth function, rather than a step
function, is large because the numbers of galaxies involved is small.

Below, we compute the density of RSCG's twice, using systems of $N \geq 3$ 
[$j=3$ in Equation~\ref{eqn:pdet}] and then using 
systems with $N \geq 4$ [$j=4$ in Equation~\ref{eqn:pdet}].  
If Equation~\ref{eqn:kappa} were strictly true for a constant $\kappa$, we
would obtain the same density for the two samples.  
However, our RSCG's with $N=3$ include both groups with a small true $\kappa$
value (these groups contain three and only three bright members)
and more distant groups with a larger true $\kappa$ value (these groups contain
more than three
bright members, but only three above
the magnitude limit).
Thus the two density estimates are estimates of two
fundamentally different quantities.
Our compact groups are not uniformly dense, and they do not
occupy the same volumes;  thus their $\kappa$ 
values actually vary.

The properties of distant
RSCG's differ from those of the nearby systems.  If the population of
galaxies in 
an RSCG brighter than $M_{\rm lim}(v)$ is in fact proportional to
$\int_{-\infty}^{M_{\rm lim}(v)} \overline{\Phi}(M) dM $ as we have
assumed, then some distant systems are far more populated, and probably 
much denser, than the nearby RSCG's.  Thus assuming a single value
of $\kappa$ for all RSCG-type systems is only an approximation.  
We do so to construct a manageable model of the selection function for
these systems.

Some assumption about the number of bright galaxies in 
compact groups is necessary in order to construct a selection function.
Monte Carlo simulations of compact groups aimed at computing a
selection function make implicit assumptions (e.g., Mendes de Oliveira
\& Hickson 1991).  Our approach enables us to make the assumptions 
explicit and their effects can be easily explored.

\begin{figure}[p]
\plottwo{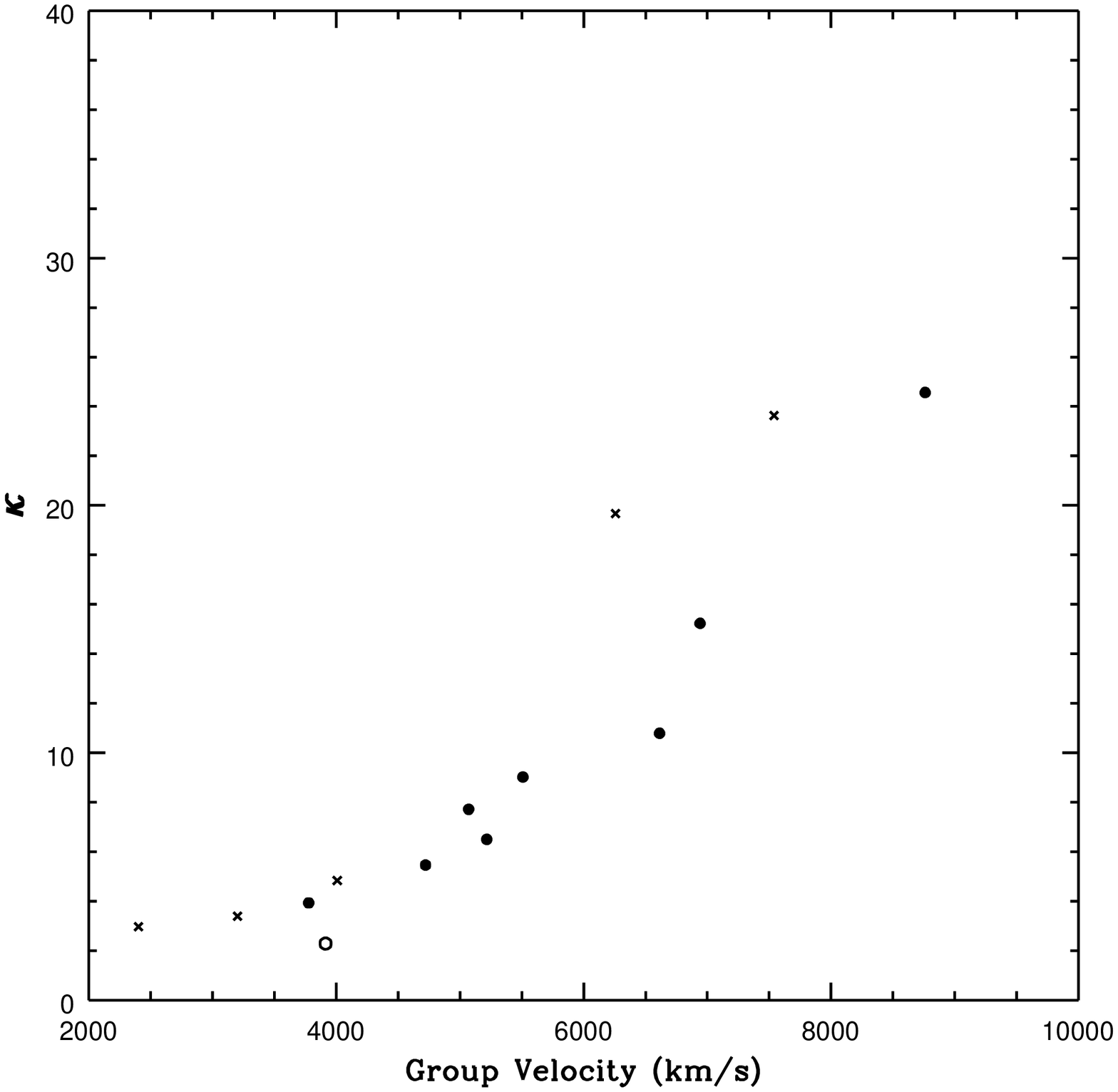}{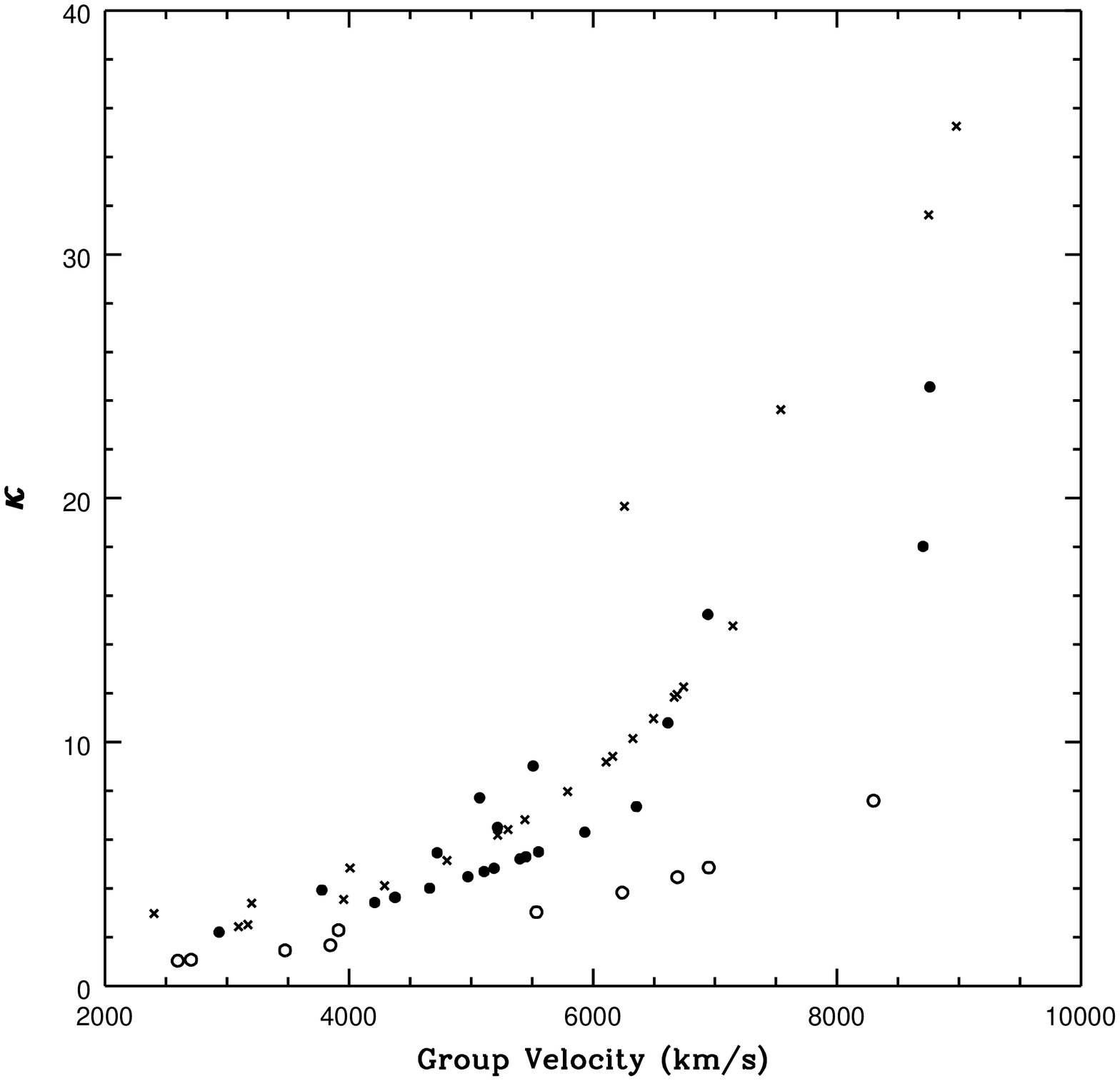}
\caption{$\kappa$ estimates vs. redshift
 for CfAnorth ($\times$), CfAsouth ($\cdot$)
and SSRS2 ($\circ$): (a) groups with $N \geq 4$, and
(b) all RSCG's.  The figure includes only groups
with median $cz < 10^{4}~{\rm km\ s}^{-1}$.  } \label{fig:k_vs_z}
\vspace{12pt}
\centerline{\epsfxsize=3.7in%
\epsffile{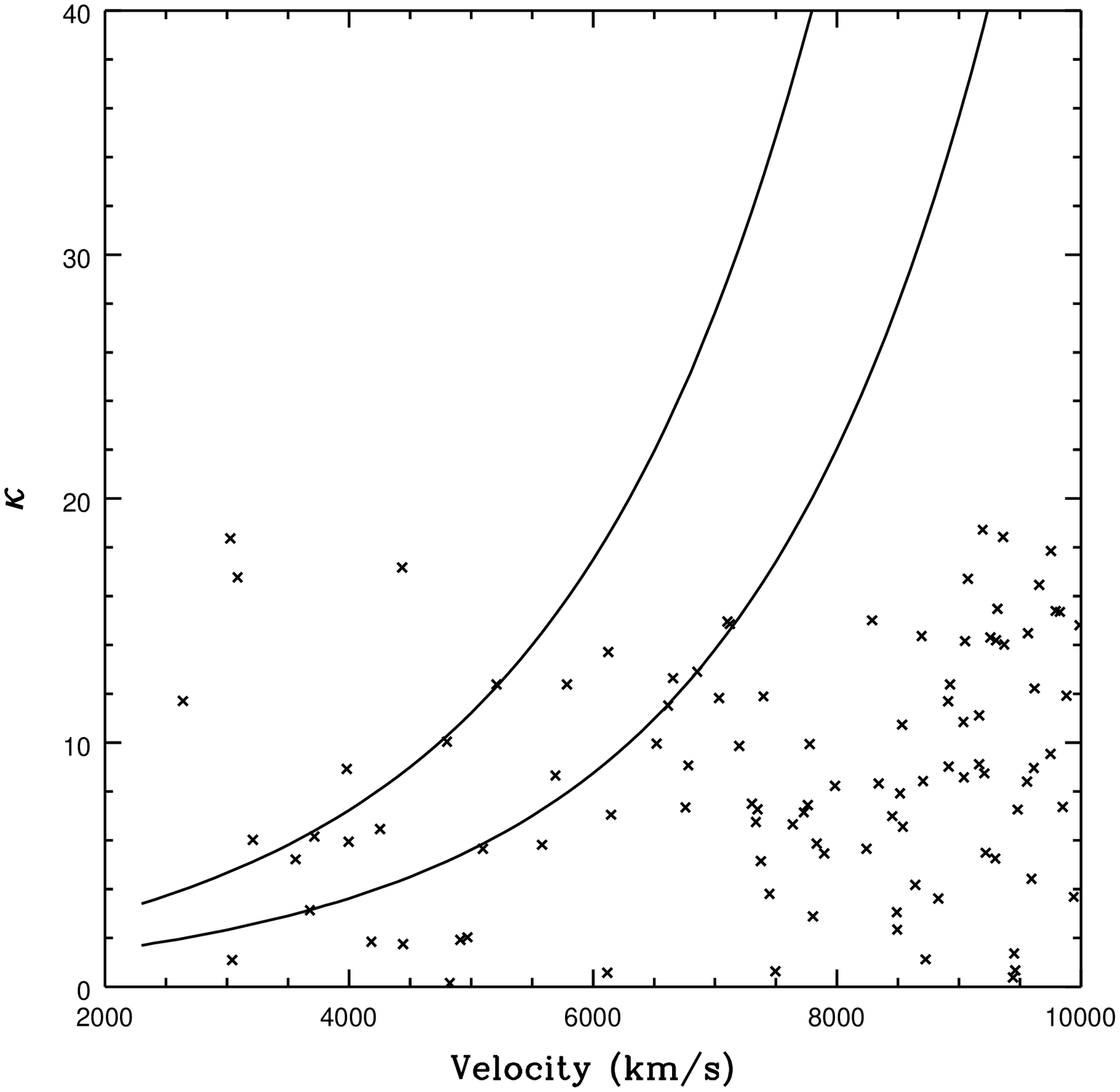}}
\caption{A random sampling of a gaussian $\kappa$ distribution.  
The lines show the  
``artificial'' limits imposed
by detecting only groups of $N \geq 3$ and
$N \leq 6$ for CfAnorth (solid line). } \label{fig:kaprange}
\end{figure}

In a magnitude-limited redshift survey, we detect only galaxies
brighter than the limiting magnitude $M_{\rm lim}(v)$.  
Because we have no information about fainter galaxies we must 
estimate $\kappa$ for each compact group;
\begin{equation}
\kappa_{\rm estim} = \left[ \frac{V_{\rm cg} \rho_{{\rm cg}, < M}}
{\int_{-\infty}^{M_{\rm lim}(v)} 
\overline{\Phi}(M') dM'} \right] 
= \left[ \frac{N}{\int_{-\infty}^{M_{\rm lim}(v)} \overline{\Phi}(M') dM'} \right].
\label{eqn:kestimate}
\end{equation}
Here, $\rho_{{\rm cg}, < M}$ is the compact group density of galaxies 
brighter than $M=M_{\rm lim}(v)$, $v$ is the median galaxy velocity 
in the RSCG, $V_{\rm cg}$ is the volume of the RSCG and $N$, as usual, is the 
number of detected galaxies in the RSCG.  
There is a lower bound and an effective upper bound 
to the actual number of observed galaxies
in a compact group. 
We select only systems with $N \geq 3$.  We expect 
from theoretical arguments (Diaferio, Geller \& Ramella \markcite{d94} 1994) 
and observation (Hickson \markcite{h82} 1982)  only systems with
$N \lesssim 6$.  
For $N \gtrsim 6$ the probability is so small that we expect none even in
a volume much larger than the region we survey.
Therefore, $N_{\rm min} =3$ and $N_{\rm max} \approx 6$.
As $N$ falls only in the
range $N_{\rm min} \leq N \lesssim N_{\rm max}$, $\kappa_{\rm estim}$ is limited
to the range $\kappa_{\rm min}(v) \leq \kappa_{\rm estim} \lesssim \kappa_{\rm max}(v)$
where
\begin{equation}
\kappa_{\rm min}(v) = \left[ \frac{3}{\int_{-\infty}^{M_{\rm lim}(v)} 
\overline{\Phi}(M') dM'} \right]
\end{equation}
and
\begin{equation}
\kappa_{\rm max}(v) = \left[ \frac{6}{\int_{-\infty}^{M_{\rm lim}(v)}
\overline{\Phi}(M') dM'} \right].
\end{equation}
Figure~\ref{fig:kaprange} shows $\kappa_{\rm min}(v)$ and $\kappa_{\rm max}(v)$
computed assuming the CfAnorth survey LF.

Figure~\ref{fig:k_vs_z} shows
$\kappa_{\rm estim}$ as a function of redshift for RSCG's with 
$cz < 10,000 {\rm~km~s}^{-1}$ in all of the
surveys; we compute $\kappa_{\rm estim}$ using 
Equation~\ref{eqn:kestimate}.  
Figure~\ref{fig:k_vs_z}a shows only the 15 groups
with $N \geq 4$;
Figure~\ref{fig:k_vs_z}b shows all of the RSCG's.
The median, first and third quartiles 
of the distribution of $\kappa$ in Figure~\ref{fig:k_vs_z}a are
7.7, 4.2, and 19, respectively; for Figure~\ref{fig:k_vs_z}b they are
5.5, 3.6, and 11, respectively.  The points are confined to a region
similar to the region between the curves in Figure~\ref{fig:kaprange}.

Figure~\ref{fig:kaprange} illustrates an important difficulty 
in {\it choosing} an appropriate $\kappa$.  Assume that 
hypothetical ``true'' compact groups have a distribution of $\kappa$
values that is, for simplicity, a gaussian with a non-negligible width.  The width
must be large enough to allow all the measured values of
$\kappa_{\rm estim}$.  Then the ``true'' distribution of
$\kappa$ for RSCG-type systems might approximate the distribution of points in
Figure~\ref{fig:kaprange}.  This figure depicts a random sampling of
a Gaussian distribution of ``$\kappa$'' values
with a peak at the median measured value of
$\kappa$ for all RSCG's and a width equal to the 
interquartile range.  The sampling is weighted according to the
volume sampled, with a minimum velocity of
2300~km~s$^{-1}$ and a maximum velocity of 10,000~km~s$^{-1}$.  
Because we use a magnitude-limited sample we measure
$\kappa$ only between $\kappa_{\rm min}(v)$ and $\kappa_{\rm max}(v)$.
Therefore, it is difficult to reconstruct the ``true'' $\kappa$ 
distribution.  We are guided only by the available estimators, the
median, first and third quartiles of the observed $\kappa_{\rm estim}$ 
distribution.

\subsubsection{Calculation of the selection function}

Assuming a constant value of $\kappa$ simplifies the
selection function.
In this case, we substitute the differential
form of Equation~\ref{eqn:kappa} into 
Equation~\ref{eqn:pdet} with $j=3$ and integrate over $\lambda_M$
to find the probability of detecting three or more galaxies
as a function of redshift,
\begin{equation}
P_{\rm detection}(v)=1-\frac{1}{2}e^{-\lambda_{M_{\rm lim}(v)}}
\left[ \lambda^2_{M_{\rm lim}(v)} + 2\lambda_{M_{\rm lim}(v)} +2 \right]. 
\label{eqn:3det}
\end{equation}
Similarly, we calculate the probability of detecting four or
more  galaxies by using $j=4$ in Equation~\ref{eqn:pdet},
\begin{equation}
P_{\rm detection}(v)=1-\frac{1}{6}e^{-\lambda_{M_{\rm lim}(v)}}
\left[ \lambda^3_{M_{\rm lim}(v)} + 3\lambda^2_{M_{\rm lim}(v)} 
+ 6\lambda_{M_{\rm lim}(v)} + 6 \right]. 
\label{eqn:4det}
\end{equation}
When computing $\lambda_{M_{\rm lim}}(v)$ we use the observed LF, which is the
true LF convolved with a gaussian of width of about the 
size of the magnitude errors (Efstathiou, Ellis \& Peterson \markcite{e88} 
1988).

\subsection{Abundance of Compact Groups}

The selection function is the first step toward calculating the volume number
density of compact groups.
In the case of a uniform spatial distribution of compact groups,
the product of $P_{\rm detection}$, the survey volume at a particular
redshift, and the average volume number density 
of compact groups ($\varrho_{\rm cga}$) is the differential expected number of
compact groups,
\begin{equation}
\frac{dN_{\rm cg}}{dv}=\varrho_{\rm cga} \frac{\Omega}{H_{0}} 
\left( \frac{v}{H_{0}} \right)^{2} P_{\rm detection}(v). \label{eqn:dndz}
\end{equation}
We average the estimated density over the
solid angle of each survey, $\Omega$.  For a survey bounded by
redshifts $v_{i}$ and $v_{f}$, the density is approximated by
\begin{equation} 
\varrho_{\rm cga,s}=\frac{N_{s}}
{\Omega \int_{v_{i}}^{v_{f}} \left( \frac{v}{H_{0}} \right)^{2} P_{\rm detection}(v) d \left( \frac{v}{H_{0}} \right) }. 
\label{eqn:bin} 
\end{equation}
$N_{s}$ is the number of compact groups with an average redshift in the
interval [$v_{i},v_{f}$].

\begin{figure}[p]
\plottwo{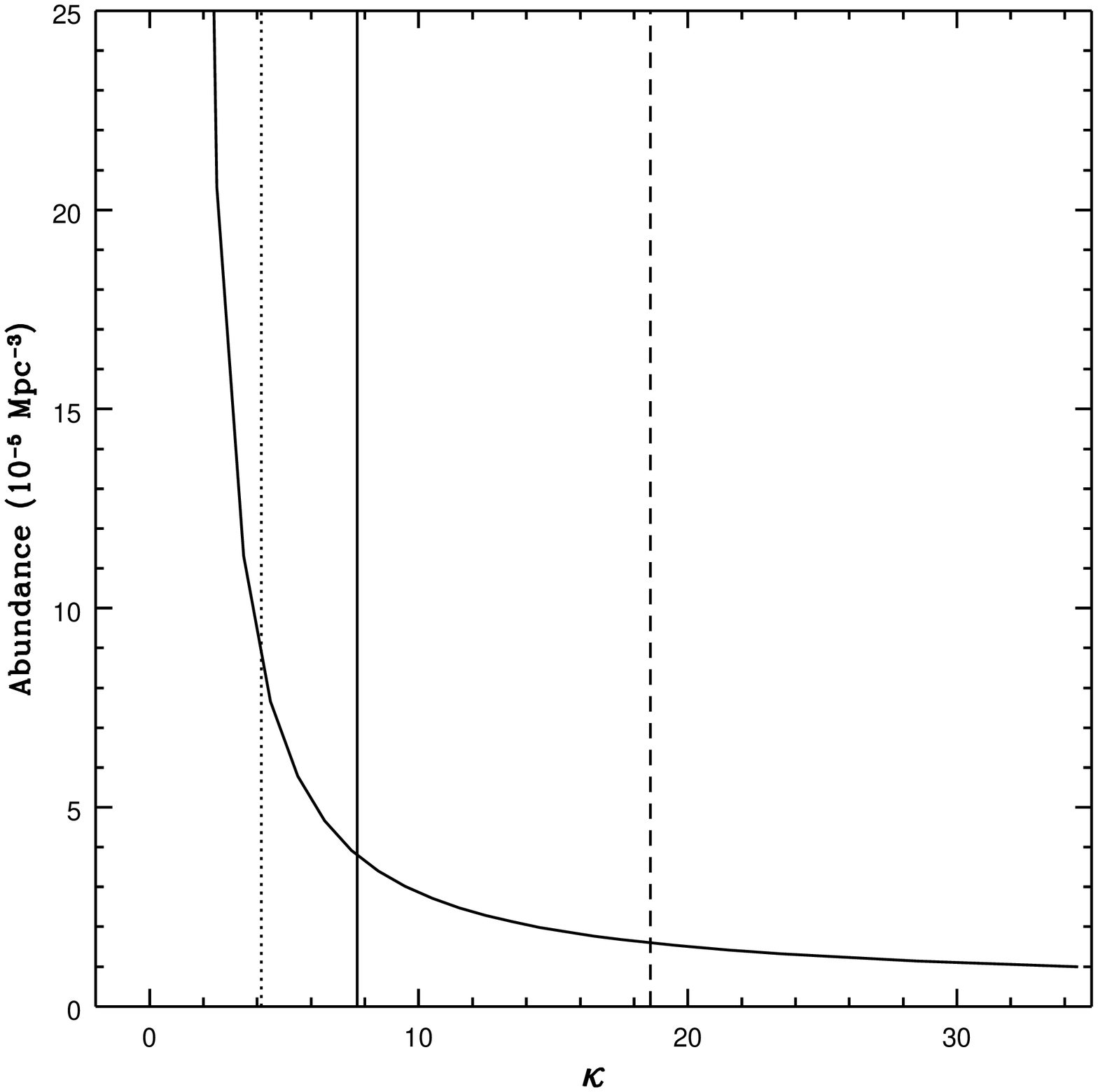}{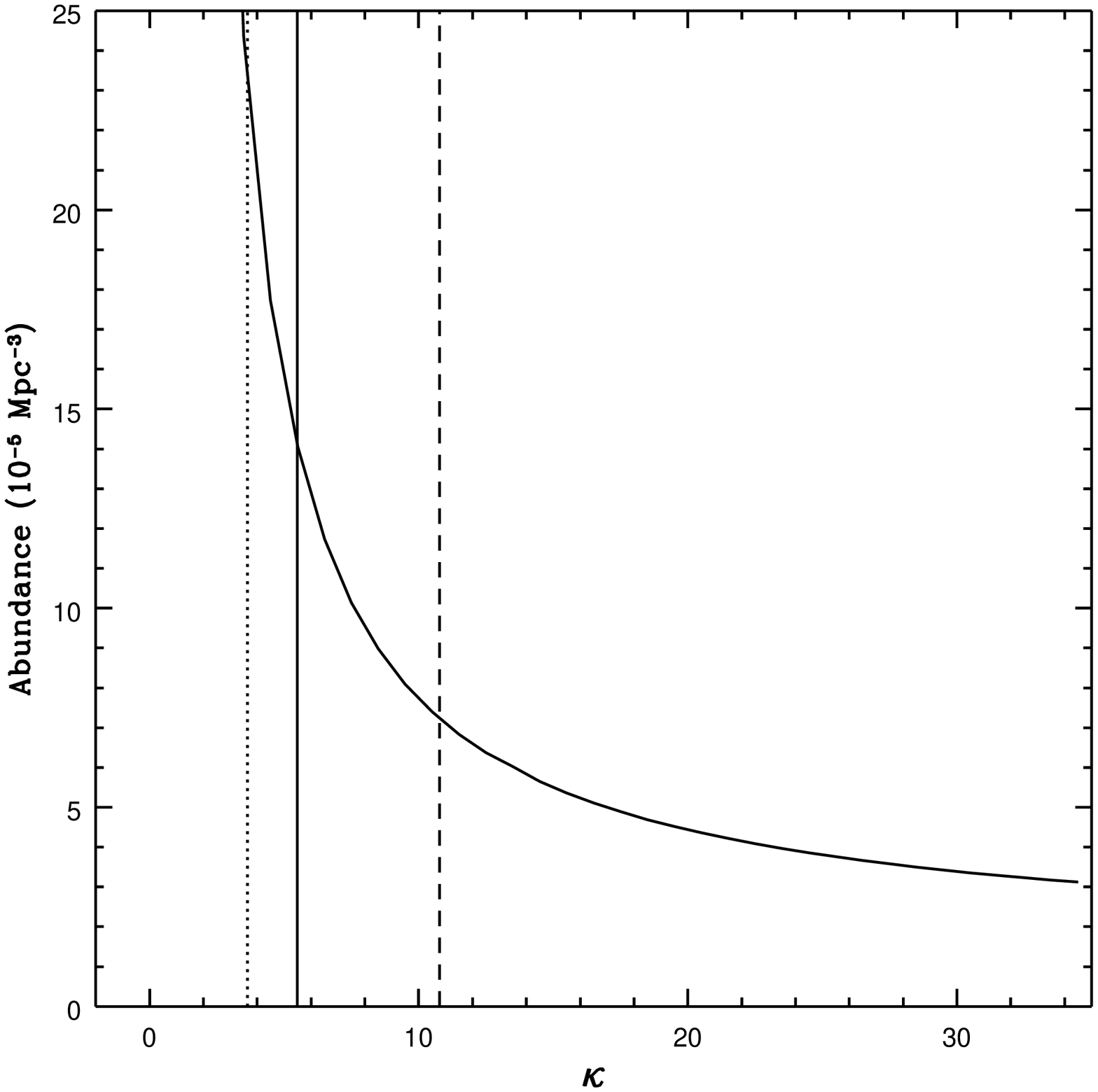}
\caption{Abundance of RSCG's as a function of $\kappa$.  
The median (solid line), first
(dotted line) and third (dashed line) quartiles of the 
$\kappa$ distribution are
shown as vertical lines:  
(a) RSCG's with $cz \geq 2300$~km~s$^{-1}$ and $N \geq 4$, and
(b) all RSCG's with $cz \geq 2300 $~km~s$^{-1}$.} \label{fig:rho_hick}
\vspace{12pt}
\centerline{\epsfxsize=3.7in%
\epsffile{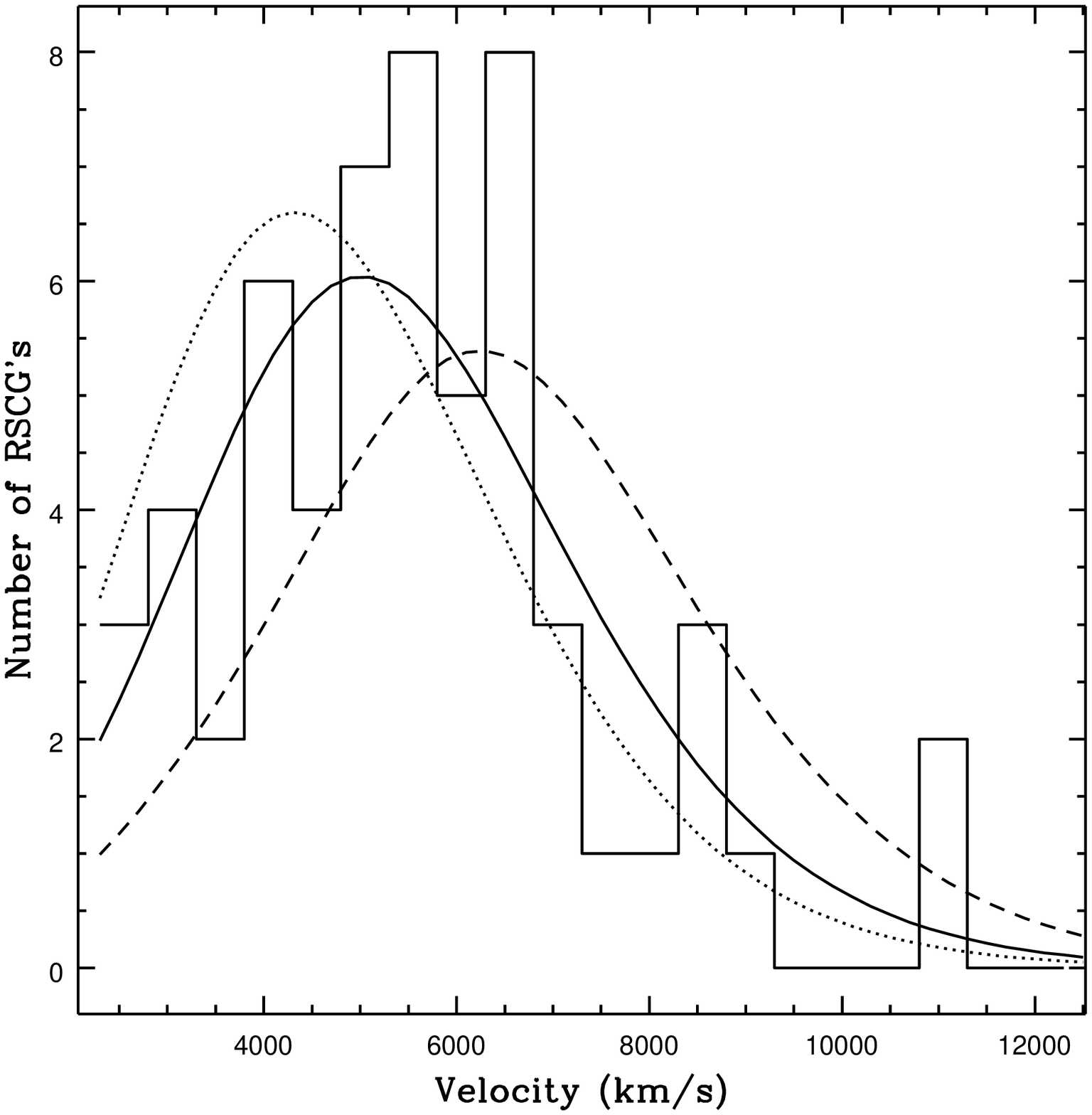}}
\caption{RSCG median velocity distribution and
models.  The three
models use the median (solid line), first quartile (dotted line), and
third quartile (dashed line) values of $\kappa$. } \label{fig:combssmod}
\end{figure}

Using Equation~\ref{eqn:bin} altered to select  groups of $N \geq 4$
($j=4$), we calculate the space density of 
RSCG's with $N \geq 4$.  There are 15 of these groups in our catalog.
Figure~\ref{fig:rho_hick}a shows $\varrho_{\rm cga}$ 
for these groups as a function of 
$\kappa$.  The figure includes data from all three surveys.
Vertical lines indicate the median, first and third quartiles of the
relevant $\kappa$ distribution.  Table~\ref{tab:rho_hick} lists the 
density of compact groups at each
of these values of $\kappa$.
The combined-survey density estimate at the median value of $\kappa$
is  $ 3.8 \times 10^{-5}$~Mpc$^{-3}$.

In selection from a redshift survey we have greater confidence in
the triplets than Hickson had for selection on the sky. 
We repeat the calculation of the space density of compact
groups using all RSCG's with average velocities greater than 
2300~km~s$^{-1}$.  We use a selection function computed
with  $j=3$ in Equation~\ref{eqn:pdet} to include the
abundant observed triplets.   
The density of these compact systems exceeds
both the original HCG abundance 
estimate of Mendes de Oliveira \& Hickson (1991) 
and our value for the density of
RSCG's derived from groups with $N \geq 4$. 
Figure~\ref{fig:rho_hick}b shows $\varrho_{\rm cga}$ for these
groups (including all surveys) as a function of $\kappa$.  Once again, 
we use vertical lines to indicate the median, first and third quartiles of
the relevant $\kappa$ distribution.
Table~\ref{tab:rho_all} lists the density of compact groups at these $\kappa$
values for each survey and for all surveys combined.  The combined-survey
density estimate at the median value of $\kappa$ is $ 1.41 \times 10^{-4}$
Mpc$^{-3}$, almost four times the estimate for $N \geq 4$.

Figure~\ref{fig:combssmod} shows our model
for the number density of all RSCG's as a function of redshift, 
$\frac{dN_{\rm cg}}{dv}$ from Equation~\ref{eqn:dndz}, 
for all the surveys combined.  
The figure shows the models for the
median (solid line), first quartile (dotted line), and third quartile
(dashed line) $\kappa$ values.  The models assume a uniform space density of
compact groups; we adopt the $\varrho_{\rm cga}$ values 
in Table~\ref{tab:rho_all}.  
We plot a histogram of the RSCG's for
comparison. The distribution corresponds roughly
to a $\kappa$ value in the proper range. However, the Figure shows 
variation in the volume number density of compact groups with redshift
which results from the large-scale structure in the nearby universe.

\begin{figure}
\plottwo{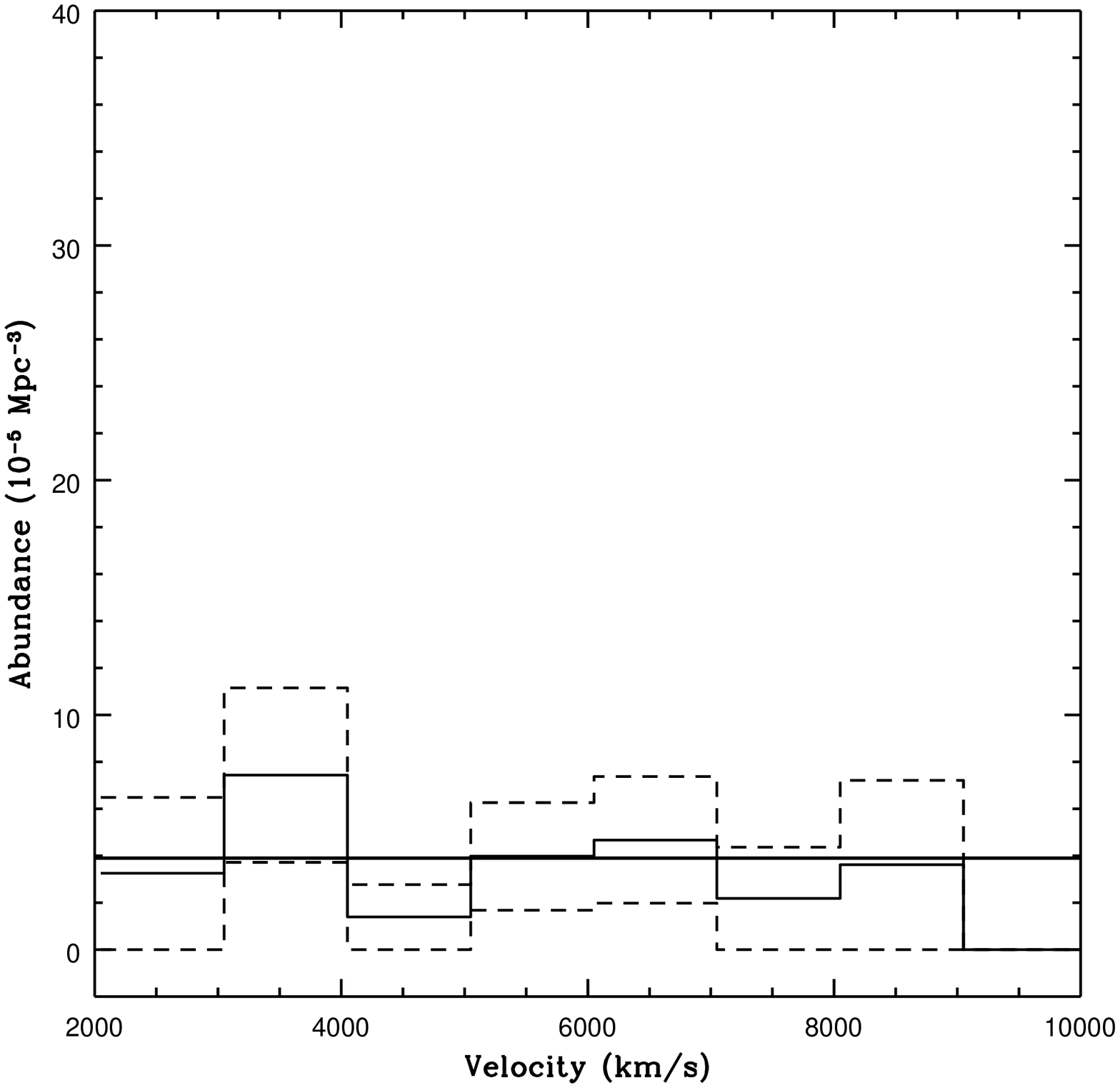}{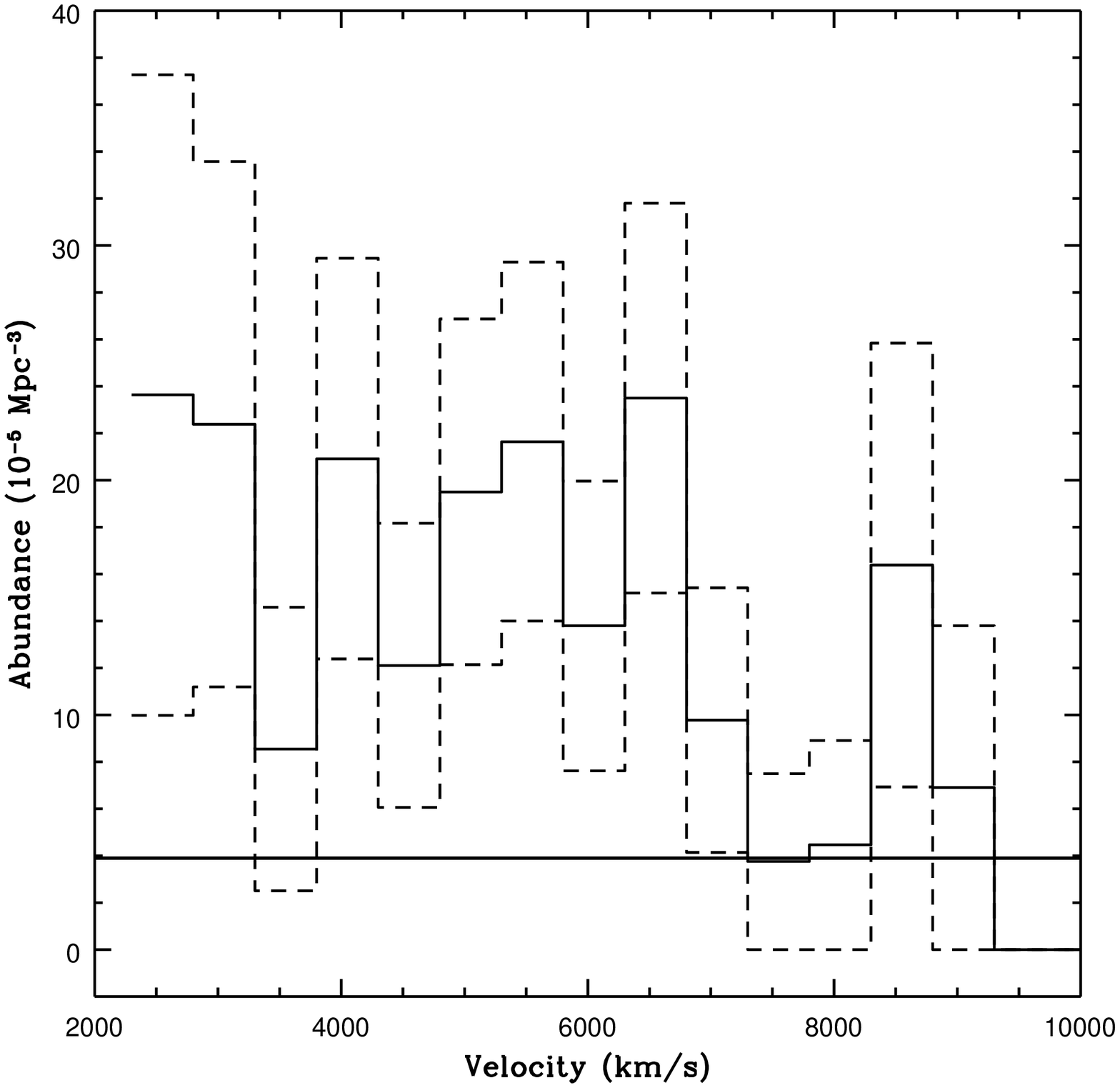}
\caption{Compact Group space density as a function of redshift computed
with the median values of $\kappa$: (a) RSCG's with $N \geq 4$, and
(b) all RSCG's. The dotted lines represent $\protect \sqrt{N}$ 
compact group population 
errors.  The horizontal line is the estimate of Mendes de Oliveira \&
Hickson (1991) of the
volume number density of HCG's.} \label{fig:nosobin}
\end{figure}

We estimate the 
volume number density of RSCG's with $N \geq 4$ and
all RSCG's again, at each redshift bin.
For this purpose, we choose a bin size of 1000~km~s$^{-1}$
for groups with $N \geq 4$ and 500~km~s$^{-1}$ for the set of all RSCG's.  
Figure~\ref{fig:nosobin} shows
$\varrho_{\rm cga}(v)$ calculated for each bin using 
Equation~\ref{eqn:bin}.  We display the results for the median value of
$\kappa$.  There is a large variation 
in RSCG density with redshift partly because of the small number
of groups.
However, Figure~\ref{fig:nosobin}b shows that the density of RSCG's
exceeds the value of Mendes de Oliveira \& Hickson (1991), the 
horizontal line, 
in most (nearby) redshift bins.

Hickson, Kindl \& Auman \markcite{h89} (1989) 
arrive at a compact group selection
function as a function of apparent magnitude empirically.  The results
are not directly comparable to our $P_{\rm detection}$ because we calculate
a selection function as a function of redshift; they calculate a 
{\it distribution} as a function of redshift.
Our density estimate derived from RSCG's with $N \geq 4$
roughly agrees with the estimated density of HCG's.
Our estimate of $\varrho_{\rm cga}$ for all compact systems,
$1.4 \times 10^{-4}\ {h}^{3}\ {\rm Mpc}^{-3}$,
greatly exceeds previous estimates of the
abundance of compact systems in the universe.  
In particular, it places stronger demands
on theories in which compact groups are
chance alignments in loose groups or ``filaments'' in the universe.

\section{Conclusion}

We apply the friends-of-friends group-finding 
algorithm to the CfA2+SSRS2 Redshift Survey to identify systems similar to
Hickson's compact groups.
The result is the first objectively defined and well-controlled sample 
of compact groups selected in redshift space.  
We evaluate the physical characteristics of the RSCG's
and find the following:

\begin{itemize}
\item The physical properties of the RSCG's (membership frequency,
velocity dispersion, density) are similar to those of the HCG's.

\item The isolation of a compact group is a 
strong function of group radius and is a poor 
predictor of the group environment; it probably has little relevance to the
dynamical history of these systems.  

\item Most RSCG's are embedded in dense environments.  Their
distribution generally follows the large-scale structure evident
in the redshift survey.  

\item The luminosity distribution of galaxies in RSCG's is
mildly inconsistent with the survey LF.  The characteristic luminosity 
is brighter and the faint end shallower for the RSCG galaxies.

\item We model the selection function of compact groups
to estimate the abundance of
RSCG's.  When we include only groups with
$N \geq 4$ the abundance is $ 3.8  \times 10^{-5}\ {h}^3\ {\rm Mpc}^{-3}$;
when we include all RSCG's the abundance is 
$1.4 \times 10^{-4}\ {h}^3\ {\rm Mpc}^{-3}$.

\end{itemize}

We plan to measure redshifts of fainter galaxies in and
around some RSCG's to explore their embedding in the surrounding environment.
In addition, we will compare the RSCG catalog to a similar objectively-defined
catalog of loose groups in the same region (Ramella et al.\ 
\markcite{r96} 1996).  We
are also searching the {\it ROSAT} all-sky survey for x-ray emission from
RSCG's.  Eventually, we intend to apply the compact group search algorithm
to a deeper redshift survey.

\acknowledgements

M.J.G., M.R. and L.daC. thank their collaborators for use of the CfA2 and
SSRS2 data in advance of publication.
We thank Mike Kurtz and Emilio Falco for providing much
assistance, and George Rybicki, James Moran, 
Antonaldo Diaferio, and Reinaldo de Carvalho 
for useful discussions and advice.
We thank the referee, Stephen Zepf, for suggestions which clarified
several important points.
E. B. is a National Science Foundation Graduate Fellow.

\begin{table}
\dummytable \label{tab:sepdist}
\end{table}

\begin{table}
\dummytable \label{tab:prop}
\end{table}

\begin{table}
\dummytable \label{tab:veldisp}
\end{table}

\begin{table}
\dummytable \label{tab:dens}
\end{table}

\begin{table}
\dummytable \label{tab:env}
\end{table}

\begin{table}
\dummytable \label{tab:lumrd}
\end{table}

\begin{table}
\dummytable \label{tab:lum}
\end{table}

\begin{table}
\dummytable \label{tab:rho_hick}
\end{table}

\begin{table}
\dummytable \label{tab:rho_all}
\end{table}

\clearpage

%FIGURES: Redshift dist. of groups:samples

\clearpage

\clearpage

\clearpage

\clearpage

%FIGURES: Confidence levels for cg LF parameters
 
\clearpage
 
% FIGURES: Kappa estimate, all 3 surveys

\clearpage

%FIGURE: Compact group space density vs. kappa
 
\clearpage 

% FIGURES: Compact group space density of redshift bins.

\begin{references}
 
\reference {a66} Arp, H. 1966, \apjs, 14, 1
\reference {b89} Barnes, J. E. 1989, \nat, 338, 123
\reference {b61} Burbidge, E. M., \& Burbidge, G. R. 1961, \aj, 66, 541
\reference {c81} Carnevali, P., Cavaliere, A., \& Santangelo, P. 1981, \apj, 249, 449
\reference {c83} Cavaliere, A., Santangelo, P., Tarquini, G., \& Vittorio, N. 1983,
in Clustering in the Universe, edited by D. Gerbal \& A. Mazure (\'{E}ditions Fronti\`{e}res, Gif-sur-Yvette), p.~25
\reference {da94} da Costa, L. N., et al. 1994, \apj, 424, 1
\reference {c94} de Carvalho, R. R., Ribeiro, A. L. B., \& Zepf, S. E. 1994, \apjs, 93, 47
\reference {ad94} Diaferio A. 1994, Ph.D. thesis, University of Milan
\reference {d94} Diaferio, A., Geller, M. J., \& Ramella, M. 1994, \aj, 107, 1623
\reference {d95} Diaferio, A., Geller, M. J., \& Ramella, M. 1995, \aj, 109, 2293
\reference {e88} Efstathiou, G., Ellis, R. S., \& Peterson, B. A. 1988, 
\mnras, 232, 431
\reference {g89} Geller, M. J., \& Huchra, J. P. 1989, Science, 246, 897
\reference {g85} Giovanelli, R., \& Haynes, M. P. 1985, \aj, 90, 2445
\reference {gh89} Giovanell, R., \& Haynes, M. P. 1989, \aj, 97, 633
\reference {g93} Giovanelli, R., \& Haynes, M. P. 1993, \aj, 105, 1271
\reference {g86} Giovanelli, R., Meyers, S. T., Roth, J., \& Haynes, M. P. 1986, \aj, 92, 250
\reference {g92} Governato, F., Bhatia, R.,  \& Chincarini, G. 1991, \apj, 371, L15
\reference {g95} Governato, F., Tozzi, P., \& Cavaliere, A. 1995, \apj, 458, 18
\reference {hay88} Haynes, M. P., Magri, C., Giovanelli, R., \& Starosta, B. M. 1988, \aj, 95, 607
\reference {hkw95} Hernquist, L., Katz, N., \& Weinberg, D. H. 1995, \apj, 442, 57\reference {h82} Hickson, P. 1982, \apj, 255, 382
\reference {h88} Hickson, P., \& Rood, H. J. 1988, \apj, 331, L69
\reference {h92} Hickson, P., Mendes de Oliveira, C., Huchra, J. P., \& Palumbo, G. G. C. 1992, \apj, 399, 353
\reference {h89} Hickson, P., Kindl, E., \& Auman, J. R. 1989, \apjs, 70, 687
\reference {hg82} Huchra, J. P., \& Geller, M. J. 1982, \apj, 257, 423
\reference {h95} Huchra, J. P., Geller, M. J., \& Corwin, Jr.,  H. G., 
1995, \apjs, 99, 391
\reference {h90} Huchra, J. P., Geller, M. J., de Lapparent, V., \& Corwin, Jr., H. G., 1990, \apjs, 72, 433
\reference {l92} Loveday, J., Peterson, B. A., Efstathiou, G., \&
Maddox, S. J. 1992, \apj, 390, 338 
\reference {m86} Mamon, G. A. 1986, \apj, 307, 426
\reference {m87} Mamon, G. A. 1987, \apj, 321, 622
\reference {m89} Mamon, G. A. 1989, \aap, 219, 98
\reference {m94} Marzke, R. O., Huchra, J. P., \& Geller, M. J. 1994, \apj, 428, 43
\reference {m91} Mendes de Oliveira, C., \& Hickson, P. 1991, \apj, 380, 30
\reference {p95} Pildis, R. A., Bregman, J. N., \& Schombert, J. M. 1995, \aj, 110, 1498
\reference {p94} Prandoni, I., Iovino, A., \& MacGillivray, H. T. 1994, \aj, 107, 1235
\reference {r89} Ramella, M., Geller, M. J., \& Huchra, J. P. 1989, \apj, 344, 57
\reference {r94} Ramella, M., Diaferio, A., Geller, M. J., \& Huchra, J. P. 1994, \aj, 107, 868
\reference {r296} Ramella, M., Pisani, A., \& Geller, M. J. 1996, \aj (submitted)
\reference {r96} Ramella, M. et al.\ 1996, in preparation
\reference {rr94} Ribeiro, A. L. B., de Carvalho, R. R., \& Zepf, S. E. 1994, \mnras, 267, L13
\reference {ro89} Rood, H. J., \& Williams, B. A. 1989, \apj, 339, 772
\reference {r77} Rose, J. A. 1977, \apj, 211, 311
\reference {r91} Rubin, V. C.,  Hunter, D. A., \& Ford, W. K. 1991, \apjs, 76, 153\reference {s76} Schechter, P. 1976, \apj, 203, 297
\reference {s73} Shakhbazyan, R. K. 1973, Astrofiz., 9, 495
\reference {v93} Vogeley, M. S. 1993, Ph.D. thesis, Harvard Univeristy
\reference {v59} Vorontsov-Vel'yaminov, B. A. 1959, Atlas and Catalogue of 
Interacting Galaxies (Sternberg Institute, Moscow State University,
Moscow), Vol. 1
\reference {v77} Vorontsov-Vel'yaminov, B. A. 1977, \aap, 28, 1
\reference {w93} Wegner, G., Haynes, M. P., \& Giovanelli, R. 1993, \aj, 105, 1251
\reference {z93} Zepf, S. E. 1993, \apj, 407, 448
 
 
 
\end{references}
\end{document}

\begin{deluxetable}{crrr}
\tablenum{1}
\tablewidth{0pc}
\tablecolumns{4}
\tablecaption{Projected separation distributions.}
\tablehead{ 
\colhead{Sample}  & \colhead{Median} & \colhead{First} & \colhead{Third}  \\
\colhead{of Groups} & & \colhead{Quartile} & \colhead{Quartile} \\
 & \colhead{(kpc)} & \colhead{(kpc)} & \colhead{(kpc)} }
\startdata
HCG's & 39 & 24 & 64 \nl
CfAnorth & 40  & 26  & 56 \nl
CfAsouth & 42  & 28  & 56 \nl
SSRS2 & 57  & 35  & 102  \nl
CfA2 + SSRS2 & 44  & 28  & 63 \nl
\enddata
\end{deluxetable}

\begin{deluxetable}{rrrrrrrc}
\tablenum{2}
\tablewidth{385pt}
\tablecolumns{8}
\tablecaption{Catalog of RSCG's.}
\tablehead{
\colhead{Group} & \colhead{$N$} & \multicolumn{2}{c}{Coordinates  (B1950)} & \colhead{$z$} & \colhead{$\theta_{G}$} & \colhead{Radius} & \colhead{Comments} \\
& &\colhead{$\alpha$} & \colhead{$\delta$} & & \colhead{(arcmin)} & \colhead{(kpc)} & }
\startdata
   1 &   5 &  00$^{h}$15$^{m}$50$^{s}$ & +29$^{\circ}$46$^{'}$33$^{''}$ & 0.023 &     1.99 &    40.20 & \nl
   2 &  3 &  00 31 18  & +06 58 30  & 0.018 &     1.57 &    24.90 & \nl
   3 &   3 &  00 36 42  & +00 36 49  & 0.014 &     1.57 &    19.20 & HCG7 \tablenotemark{a}\nl
   4 &   3 &  00 40 21  & -23 49 37  & 0.022 &     1.02 &    19.90 & \nl
\tablenotetext{a}{We label RSCG's as HCG's if the RSCG contains all or part of the HCG.}
   5 &   5 &  01 04 38  & +32 05 23  & 0.017 &     3.40 &    50.20 & \nl
   6 &   3 &  01 13 15  & +46 28 30  & 0.017 &     1.63 &    24.60 & \nl
   7 &   3 &  01 20 20  & +33 11 32  & 0.015 &     2.94 &    37.40 & \nl
   8 &   5 &  01 20 41  & +33 00 39  & 0.017 &     3.66 &    55.60 & \nl
   9 &   3 &  01 22 36  & +14 36 30  & 0.021 &     1.54 &    28.40 & \nl
  10 &   3 &  01 22 53  & -01 46 42  & 0.017 &     1.90 &    28.20 & \nl
  11 &   5 &  01 23 21  & -01 35 11  & 0.018 &     2.14 &    34.30 & \nl
  12 &   3 &  01 23 34  & +34 27 18  & 0.016 &     2.67 &    36.20 & HCG 10\nl
  13 & 3 &  01 40 55  & -34 27 43  & 0.013 &     1.90 &    21.20 & \nl
  14 &   3 &  01 46 56  & +21 45 29  & 0.010 &     3.83 &    32.70 & \nl
  15 &   4 &  01 49 45  & +35 54 00  & 0.016 &     2.09 &    28.70 & \nl
  16 &   3 &  01 52 03  & +36 39 30  & 0.018 &     1.63 &    25.50 & \nl
  17 &   3 &  01 53 45  & +05 23 58  & 0.019 &     1.23 &    19.80 & \nl
  18 &   3 &  01 53 53  & +32 47 17  & 0.015 &     1.72 &    21.90 & \nl
  19 &   4 &  02 07 04  & -10 23 40  & 0.013 &     3.11 &    35.30 & HCG 16 \nl
  20 &   3 &  03 01 12  & -15 49 55  & 0.009 &     2.06 &    16.20 & HCG 22\nl
  21 &   3 &  03 16 18  & +41 22 48  & 0.017 &     2.47 &    35.70 & \nl
  22 &   3 &  03 24 16  & -21 27 35  & 0.005 &     4.18 &    18.80 & \nl
  23 &   3 &  03 33 16  & -32 43 04  & 0.005 &     6.45 &    26.90 & \nl
  24 &  13 &  03 34 38  & -35 41 22  & 0.005 &    28.56 &   113.40 & \nl
  25 &   4 &  03 37 49  & -18 45 45  & 0.005 &     9.53 &    44.30 & \nl
  26 &   4 &  04 10 30  & -33 03 12  & 0.004 &     7.07 &    24.50 & \nl
  27 &   3 &  04 13 10  & -28 36 29  & 0.018 &     0.30 &     4.80 & \nl
  28 &   3 &  09 01 09  & +22 10 30  & 0.010 &     4.55 &    41.00 & \nl
  29 &   4 &  09 07 13  & +23 03 02  & 0.038 &     1.00 &    32.80 & \nl
  30 &   3 &  09 09 30  & +35 11 28  & 0.007 &     4.04 &    23.70 & \nl
  31 &   3 &  09 14 09  & +42 10 00  & 0.006 &     3.05 &    15.30 & \nl
  32 &   3 &  09 16 42  & +33 57 18  & 0.020 &     0.72 &    12.80 & \nl
  33 &   3 &  09 31 12  & +10 18 30  & 0.011 &     3.80 &    35.00 & \nl
  34 &   3 &  09 40 15  & +32 08 32  & 0.005 &     5.54 &    24.90 & \nl
  35 &   4 &  10 15 16  & +22 02 28  & 0.004 &     8.16 &    31.70 & HCG 44\nl
  36 &   3 &  10 45 30  & +12 49 32  & 0.003 &     4.72 &    12.60 & \nl
  37 &   3 &  10 45 31  & +14 22 00  & 0.002 &    11.60 &    21.50 & \nl
  38 &   3 &  10 49 00  & +33 07 27  & 0.005 &     7.46 &    30.80 & \nl
  39 &   4 &  10 52 27  & +17 26 30  & 0.004 &    14.03 &    45.20 & \nl
  40 &   5 &  11 13 38  & +18 15 34  & 0.003 &    14.09 &    39.00 & \nl
  41 &   4 &  11 24 38  & +17 17 37  & 0.004 &    14.20 &    47.90 & \nl
  42 &   3 &  11 34 15  & +20 15 55  & 0.021 &     1.27 &    23.40 & \nl
  43 &   3 &  11 35 17  & +22 15 24  & 0.030 &     0.46 &    12.00 & HCG 57\nl
  44 &   6 &  11 41 25  & +20 13 23  & 0.021 &     3.20 &    58.20 & \nl
  45 &   3 &  11 41 44  & +20 06 46  & 0.018 &     0.62 &     9.80 & \nl
  46&   4 &  11 45 52  & +12 59 21  & 0.013 &     1.53 &    17.90 & HCG 59\nl
  47 &   4 &  11 55 03  & +32 34 60  & 0.011 &     2.14 &    20.00 & \nl
  48 &   3 &  12 04 18  & +43 24 30  & 0.003 &     9.82 &    23.00 & \nl
  49 &   3 &  12 09 51  & +29 26 43  & 0.013 &     1.87 &    21.60 & HCG 61\nl
  50 &   3 &  12 12 27  & +13 25 00  & 0.002 &    20.64 &    37.70 & \nl
  51 &   7 &  12 18 30  & +29 56 45  & 0.002 &    26.01 &    49.70 & \nl
  52 &   3 &  12 18 30  & +40 09 30  & 0.022 &     1.50 &    29.10 & \nl
  53 &   3 &  12 19 24  & +14 57 30  & 0.004 &     7.34 &    24.20 & \nl
  54 &   3 &  12 22 51  & +18 28 30  & 0.003 &     7.84 &    17.30 & \nl
  55 &   4 &  12 24 04  & +09 18 39  & 0.025 &     2.15 &    47.10 & \nl
  56 &   5 &  12 24 56  & +12 41 59  & 0.003 &    10.05 &    26.70 & \nl
  57 &   6 &  12 25 54  & +09 55 58  & 0.002 &    24.08 &    44.80 & \nl
  58 &   3 &  12 25 56  & +13 21 59  & 0.002 &    11.68 &    23.20 & \nl
  59 &   3 &  12 26 56  & +14 17 51  & 0.004 &     6.94 &    24.50 & \nl
  60 &   4 &  12 26 58  & +09 08 25  & 0.002 &    19.53 &    40.90 & \nl
  61 &   5 &  12 28 00  & +12 38 11  & 0.005 &     7.81 &    31.10 & \nl 
  62 &   5 &  12 28 57  & +11 57 57  & 0.003 &    14.46 &    43.70 & \nl
  63 &   4 &  12 32 58  & +12 45 45  & 0.002 &    15.76 &    25.50 & \nl
  64 &   3 &  12 38 60  & +26 19 30  & 0.016 &     1.50 &    21.00 & \nl
  65 &   7 &  12 39 48  & +11 38 32  & 0.004 &    22.67 &    72.20 & \nl
  66 &   3 &  12 40 47  & +13 28 12  & 0.003 &     9.91 &    30.10 & \nl
  67 &   3 &  12 57 08  & +28 13 37  & 0.024 &     0.64 &    13.40 & \nl
  68 &   3 &  12 57 46  & +28 14 26  & 0.022 &     1.76 &    33.30 & \nl
  69 &   3 &  13 14 57  & +20 53 13  & 0.022 &     1.37 &    26.80 & \nl
  70 &   3 &  13 22 42  & +36 40 00  & 0.018 &     2.33 &    36.00 & HCG 68\nl
  71 &   5 &  13 51 34  & +40 33 50  & 0.008 &     4.52 &    31.60 & \nl
  72 &   3 &  13 54 45  & +12 15 00  & 0.021 &     1.24 &    22.20 & \nl
  73 &   3 &  14 00 18  & +09 36 00  & 0.019 &     1.48 &    24.90 & \nl
  74 &   3 &  14 09 54  & +16 05 30  & 0.017 &     1.50 &    22.80 & \nl
  75 &   3 &  14 44 42  & +11 48 00  & 0.029 &     1.47 &    37.40 & \nl
  76 &   3 &  15 04 30  & +13 03 00  & 0.022 &     1.46 &    28.40 & \nl
  77 &   3 &  15 57 00  & +20 53 59  & 0.014 &     0.30 &     3.80 & HCG 79\nl
  78 &   4 &  20 44 39  & +00 10 60  & 0.013 &     4.59 &    50.40 & \nl
  79 &   3 &  21 08 48  & -23 22 05  & 0.036 &     0.64 &    20.30 & \nl
  80 &   3 &  21 59 10  & -32 10 19  & 0.009 &     3.67 &    27.70 & HCG 90\nl
  81 &   3 &  22 06 20  & -28 02 30  & 0.023 &     1.09 &    22.00 & HCG 91\nl
  82 &   4 &  22 33 40  & +33 42 02  & 0.022 &     1.52 &    29.10 & HCG 92\nl
  83 &   3 &  22 59 28  & +15 42 24  & 0.007 &     1.13 &     6.40 & \nl
  84 &   3 &  23 11 51  & -02 57 31  & 0.012 &     2.64 &    26.60 & \nl
  85 &   3 &  23 19 09  & +26 49 30  & 0.020 &     1.64 &    28.30 & \nl
  86 &   4 &  23 36 04  & +26 44 47  & 0.029 &     1.26 &    32.10 & \nl
  87 &   3 &  23 44 49  & -02 36 14  & 0.021 &     1.52 &    27.60 & HCG 97\nl
  88 &   3 &  23 44 53  & -28 24 02  & 0.028 &     0.79 &    19.10 & \nl
  89 &   3 &  23 58 09  & +28 06 55  & 0.029 &     1.11 &    28.00 & HCG 99\nl
\tablecomments{Catalog of RSCG's: (1)~Group number, (2)~Number of galaxies in group, (3)~Right ascension, (4)~Declination, (5)~Redshift, (6)~Angular radius, (7)~Linear radius, and (8)~Comments.}
\enddata
\end{deluxetable}

\begin{deluxetable}{crrrr}
\tablenum{3}
\tablewidth{0pc}
\tablecolumns{5}
\tablecaption{Compact group velocity dispersion distributions.}
\tablehead{
\colhead{Sample}  & \colhead{Number} & \colhead{Median} & \colhead{First}  & \colhead{Third}  \\
\colhead{of Groups} & \colhead{of Groups} & & \colhead{Quartile} & \colhead{Quartile} \\
& & \colhead{(km~s$^{-1}$)} & \colhead{(km~s$^{-1}$)} & \colhead{(km~s$^{-1}$)} }
\startdata
 HCG's & 23 &123 & 70 & 291  \nl
($N=3$) & & & & \nl
& & & & \nl
RSCG's & 59 & 129 & 68 & 249  \nl
($N=3$) & & & & \nl
& & & & \nl
HCG's & 69 & 214 & 148 & 301 \nl
($N \geq 4$) & & & & \nl
& & & & \nl
RSCG's & 30 & 262 & 154 & 376 \nl
($N \geq 4$) & & & & \nl
& & & & \nl
\enddata
\end{deluxetable}

\begin{deluxetable}{crrrr}
\tablenum{4}
\tablewidth{0pc}
\tablecolumns{5}
\tablecaption{Compact group density distributions.}
\tablehead{
\colhead{Sample}  & \colhead{Number} & \colhead{Median} & \colhead{First}  & \colhead{Third}  \\
\colhead{of Groups} & \colhead{of Groups} & & \colhead{Quartile} & \colhead{Quartile} \\
 & & \colhead{($10^{4}$ $h^{3}$ Mpc$^{-3}$)} 
& \colhead{($10^{4}$ $h^{3}$ Mpc$^{-3}$)} 
& \colhead{($10^{4}$ $h^{3}$ Mpc$^{-3}$)}  }
\startdata
HCG's & 23 & 3.6 & 1.0  & 7.9  \nl
($N=3$) & & & & \nl
& & & & \nl
RSCG's & 59 & 5.1  & 3.1 & 9.2 \nl
($N=3$) & & & & \nl
& & & & \nl
HCG's & 69 & 2.4 & 0.76  & 5.1 \nl
($N \geq 4$) & & & & \nl
& & & & \nl
RSCG's & 30  & 1.9  & 0.94  & 3.9  \nl
($N \geq 4$) & & & & \nl
& & & & \nl
\enddata
\end{deluxetable}

\begin{deluxetable}{rrrrrrrrc}
\tablenum{5}
\tablewidth{0pc}
\tablecolumns{9}
\tablecaption{Environments of RSCG's.}
\tablehead{
\colhead{Group} & \colhead{$N$} & \colhead{$v_{\rm med}$} & \colhead{$N_{n}$} & \colhead{$N_{\rm int,env}$} & \colhead{$\frac{\langle \rho_{\rm env} \rangle }{\langle \overline{\rho} \rangle}$} & \colhead{$N_{\rm int,cg}$} & \colhead{$\frac{\langle \rho_{\rm cg} \rangle }{\langle \overline{\rho} \rangle }$} & \colhead{Comments} \\ 
 & & \colhead{(km s$^{-1}$)} & & & & \colhead{($\times 10^{-3}$)} & \colhead{($\times 10^2$)}& }
\startdata
1&    5&  6942.&  18&    1.3&   13.8& 1.1&  45.&                    \nl
2&    3&  5449.&   6&    2.3&    2.6&0.4&  69.&                    \nl
3&    3&  4210.&  14&    3.5&    4.0&0.4&  71.&                    \nl
4&    3&  6692.&   4&    2.0&    2.0& 0.3& 119.&                    \nl
5&    5&  5071.&  48&    2.6&   18.4& 3.1&  15.&                    \nl
6&    3&  5188.&   3&    2.5&    1.2& 0.5&  57.&near survey edge    \nl
7&    3&  4375.&  59&    3.3&   17.6& 1.9&  15.&                    \nl
8&    4&  5217.&  58&    2.5&   23.4& 4.1&   9.&                    \nl
9&    3&  6356.&   7&    1.6&    4.3& 0.4&  71.&                    \nl
10 &  3&  5107.&  40&    2.6&   15.5& 0.8&  36.&near survey edge    \nl
11 &    5&  5509.&  41&    2.2&   18.4& 0.9&  57.&near survey edge    \nl
12 &  3&  4660.&  45&    3.0&   14.9& 1.6&  19.&                    \nl
13 &    3&  3846.&   8&    5.4&    1.5& 0.9&  32.&                    \nl
14 &    3&  2935.&  11&    5.5&    2.0& 1.9&  15.&                    \nl
15 &    4&  4722.&  65&    3.0&   22.0& 1.0&  41.&                    \nl
16 &    3&  5400.&  54&    2.3&   23.3& 0.6&  51.&                    \nl
17 &    3&  5553.&  15&    2.2&    6.8& 0.4&  81.&                    \nl
18 &    3&  4379.&  18&    3.3&    5.4& 0.6&  53.&                    \nl
19 &    4&  3914.&   9&    5.3&    1.7& 2.3&  17.&                    \nl
20 &    3&  2705.&  15&    8.5&    1.8& 0.7&  42.&                    \nl
21 &    3&  4974.&  39&    2.7&   14.4& 1.5&  20.&                    \nl
27 &    3&  5536.&   3&    3.0&    1.0& 0.0&1500.&                    \nl
28 &    3&  3093.&   8&   12.2&    0.7& 6.6&   4.&                    \nl
29 &    4& 11252.&   6&    0.2&   29.7& 0.1& 444.&                    \nl
32 &    3&  6106.&  10&    3.2&    3.1& 0.3& 103.&                    \nl
33 &    3&  3171.&  14&   11.8&    1.2& 4.7&   6.&near survey edge    \nl
42 &    3&  6328.&  16&    2.9&    5.6& 0.5&  54.&                    \nl
43 &    3&  8977.&   8&    0.8&   10.6& 0.0& 750.&                    \nl
44 &    6&  6258.&  61&    3.0&   20.5& 4.6&  13.&                    \nl
45 &    3&  5440.&  49&    4.3&   11.3& 0.2& 176.&                    \nl
46 &    4&  4008.&  12&    8.2&    1.5& 1.1&  37.&                    \nl
47 &    4&  3200.&  20&   11.7&    1.7& 1.6&  25.&                    \nl
49 &    3&  3956.&  18&    8.4&    2.1& 1.3&  23.&                    \nl
52 &    3&  6665.&   5&    2.5&    2.0& 0.7&  42.&                    \nl
55 &    4&  7538.&  14&    1.6&    8.8& 1.5&  26.&near survey edge    \nl
64 &    3&  4801.&   7&    5.8&    1.2& 0.9&  34.&                    \nl
67 &    3&  7145.&  97&    1.9&   50.0& 0.1& 250.&                    \nl
68 &    3&  6497.&  96&    2.7&   36.1& 1.2&  26.&                    \nl
69 &    3&  6743.&   3&    2.4&    1.3& 0.5&  55.&                    \nl
70 &    3&  5303.&   9&    4.6&    1.9& 2.2&  13.&                    \nl
71 &    5&  2401.&  45&   16.7&    2.7& 6.2&   8.&                    \nl
72 &    3&  6160.&   9&    3.1&    2.9& 0.5&  62.&                    \nl
73 &    3&  5792.&  32&    3.7&    8.7& 0.8&  36.&near survey edge    \nl
74 &    3&  5219.&  15&    4.8&    3.1& 0.8&  39.&                    \nl
75 &    3&  8753.&  14&    0.8&   16.5& 0.4&  73.&                    \nl
76 &    3&  6688.&  10&    2.4&    4.1& 0.7&  45.&                    \nl
77 &    3&  4292.&  21&    7.2&    2.9& 0.0& 750.&                    \nl
78 &    4&  3776.&  11&    4.1&    2.7& 4.1&   9.&                    \nl
79 &    3& 10877.&   8&    0.5&   17.2& 0.1& 428.&                    \nl
80 &    3&  2596.&  27&    8.9&    3.0& 2.2&  13.&                    \nl
81 &    3&  6949.&  12&    1.9&    6.5& 0.3&  93.&                    \nl
82 &    4&  6614.&  16&    1.5&   10.8& 0.6&  70.&                    \nl
84 &    3&  3475.&   5&    6.2&    0.8& 1.4&  21.&near survey edge    \nl
85 &    3&  5932.&   9&    1.9&    4.7& 0.5&  54.&                    \nl
86 &    4&  8760.&  20&    0.6&   32.4& 0.3& 129.&                    \nl
87 &    3&  6239.&   6&    2.4&    2.5& 0.8&  38.&near survey edge    \nl
88 &    3&  8299.&  18&    1.2&   15.4& 0.1& 230.&                    \nl
89 &    3&  8705.&   5&    0.6&    7.9& 0.2& 150.&                    \nl
\tablecomments{Environments of RSCG's: (1) Group number, 
(2) number of galaxies in group, (3) median group velocity, 
(4) number of galaxies in neighborhood, (5) expected number 
of interlopers in neighborhood, (6) resulting ratio of environment density to 
average survey density, (7) expected number of interlopers in compact 
group volume, and (8) resulting ratio of compact group density to average density.}
\enddata
\end{deluxetable}

\begin{deluxetable}{crr}
\tablenum{6}
\tablewidth{0pc}
\tablecolumns{3}
\tablecaption{Survey schechter parameters.}
\tablehead{
\colhead{Survey} & \colhead{$\alpha_{\rm survey}$} & \colhead{$M_{\star,{\rm survey}}$} }
\startdata
CfAnorth \tablenotemark{a} & -1.03 & -18.67 \nl
CfAsouth \tablenotemark{a} & -0.89 & -18.93 \nl
CfA-Combined \tablenotemark{a} & -1.00 & -18.75 \nl
SSRS2 \tablenotemark{b} & -1.20 & -19.50 \nl
\tablenotetext{a} {Marzke, Huchra \& Geller (1994).}
\tablenotetext{b} {da Costa et al. (1994).}
\enddata
\end{deluxetable}

\begin{deluxetable}{crr}
\tablenum{7}
\tablewidth{0pc}
\tablecolumns{3}
\tablecaption{Best-Fit RSCG Schechter parameters.}
\tablehead{
\colhead{Survey} & \colhead{$\alpha_{\rm cg}$} & \colhead{$M_{\star,{\rm cg}}$} }
\startdata
CfAnorth & -1.01 & -19.33 \nl
CfAsouth & -0.56 & -19.29 \nl
SSRS2 & -0.79 & -19.92 \nl
\enddata
\end{deluxetable}

\begin{deluxetable}{crrr}
\tablenum{8}
\tablewidth{0pc}
\tablecolumns{4}
\tablecaption{Abundances of RSCG's with $N \geq 4$.}
\tablehead{
\colhead{Survey} & \multicolumn{3}{c}{Compact Group Abundance} \\
	\cline{2-4} \\
       & \multicolumn{3}{c}{$(10^{-5}$ Mpc$^{-3})$} \\ 
       & \multicolumn{1}{c}{at Median}  & \multicolumn{1}{c}{at 1st} & \multicolumn{1}{c}{at 3rd} \\ 
       & \colhead{$\kappa$} & \colhead{Quartile $\kappa$} & \colhead{Quartile $\kappa$} }
\startdata
Cfanorth & 8.4  & 20 & 3.4 \nl
Cfasouth & 6.6  & 16  & 2.6 \nl
SSRS2 & 0.50  & 1.0  & 0.22 \nl
All Combined & 3.8  & 8.6  & 1.6\nl
\enddata
\end{deluxetable}

\begin{deluxetable}{crrr}
\tablenum{9}
\tablewidth{0pc}
\tablecolumns{4}
\tablecaption{Abundances of RSCG's with $N =3$.}
\tablehead{
\colhead{Survey} & \multicolumn{3}{c}{Compact Group Abundance} \\
        \cline{2-4} \\
       & \multicolumn{3}{c}{$(10^{-5}$ Mpc$^{-3})$} \\ 
       & \multicolumn{1}{c}{at Median}  & \multicolumn{1}{c}{at 1st} & \multicolumn{1}{c}{at 3rd} \\ 
       & \colhead{$\kappa$} & \colhead{Quartile $\kappa$} & \colhead{Quartile $\kappa$} }
\startdata
Cfanorth & 34  & 57  & 17  \nl
Cfasouth & 17 & 30 & 8.4 \nl
SSRS2 & 5.3  & 8.3  & 2.8 \nl
All Combined & 14  & 23  & 7.2 \nl
\enddata
\end{deluxetable}